\documentclass[twocolumn,aps,prb,superscriptaddress,10pt]{revtex4-2}

\usepackage{graphicx}
\usepackage{amsmath}
\usepackage{mathptmx}
\usepackage{siunitx}
\usepackage{units}

\begin{document}

\title{Spectrally stable nitrogen-vacancy centers in diamond formed by carbon implantation into thin microstructures}
\author{V. Yurgens}
\author{A. Corazza}
\author{J. A. Zuber}
\author{M. Gruet}
\author{M. Kasperczyk}
\author{B. J. Shields}
\author{R. J. Warburton}
\author{Y. Fontana}
\author{P. Maletinsky*}
\email{patrick.maletinsky@unibas.ch}
\affiliation{Department of Physics, University of Basel, CH-4056 Basel, Switzerland}


\begin{abstract}
The nitrogen-vacancy center (NV) in diamond, with its exceptional spin coherence and convenience in optical spin initialization and readout, is increasingly used both as a quantum sensor and as a building block for quantum networks. Employing photonic structures for maximizing the photon collection efficiency in these applications typically leads to broadened optical linewidths for the emitters, which are commonly created via nitrogen ion implantation. With studies showing that only native nitrogen atoms contribute to optically coherent NVs, a natural conclusion is to either avoid implantation completely, or substitute nitrogen implantation by an alternative approach to vacancy creation. Here, we demonstrate that implantation of carbon ions yields a comparable density of NVs as implantation of nitrogen ions, and that it results in NV populations with narrow optical linewidths and low charge-noise levels even in thin diamond microstructures. We measure a median NV linewidth of \unit[150]{MHz} for structures thinner than \unit[5]{\textmu m}, with no trend of increasing linewidths down to the thinnest measured structure of \unit[1.9]{\textmu m}. We propose a modified NV creation procedure in which the implantation is carried out after instead of before the diamond fabrication processes, and confirm our results in multiple samples implanted with different ion energies and fluences.
\end{abstract}

\maketitle


The negatively charged nitrogen vacancy center (NV) in diamond is a widely applicable system for both quantum communication~\cite{Awschalom2018} and quantum sensing~\cite{Degen2017}. It features an optically addressable electron spin with long spin coherence both at low- and at room temperature~\cite{Balasubramanian2009,Bar-Gill2013,Abobeih2018}, hyperfine-coupled nuclear spins that can be used as long-term storage qubits~\cite{Bradley2019}, and an electronic structure that enables sensing of magnetic, electric or mechanical fields as well as temperature~\cite{Doherty2013}. Nanoscale probing of magnetism in 2D materials~\cite{Thiel2019}, imaging of domain walls in antiferromagnets~\cite{Hedrich2021}, and measurement of current flow in graphene~\cite{Palm2022} have benefited from this sensitivity, while demonstrations of entanglement between distant NVs~\cite{Hensen2015} and implementations of few-node quantum networks~\cite{Pompili2021} build on the NV's electronic, nuclear, as well as optical coherence. Because these applications rely on optical initialization and interrogation, an efficient optical interface is critical. For NVs deep in the bulk, the solution has so far been to employ solid immersion lenses~\cite{Jamali2014,Hadden2010}, which improve the collection efficiency by up to a factor of 10 over unstructured diamond but do not increase the weak NV zero-phonon line (ZPL) fraction. Approaches based on optical resonators such as open microcavities~\cite{Barbour2011b,Albrecht2013,Johnson2015,Kaupp2016,Janitz2015,Bogdanovic2017,Riedel2017c} shorten the radiative lifetime, increase the fraction of photons emitted into the ZPL, and improve the collection efficiency, but require more extensive diamond microfabrication. For quantum sensing, the common approach to increase the collection efficiency is to use waveguiding photonic structures as scanning probes~\cite{Rohner2019,Hedrich2020}, which also requires a considerable amount of diamond microstructuring.

\begin{figure}
    \includegraphics{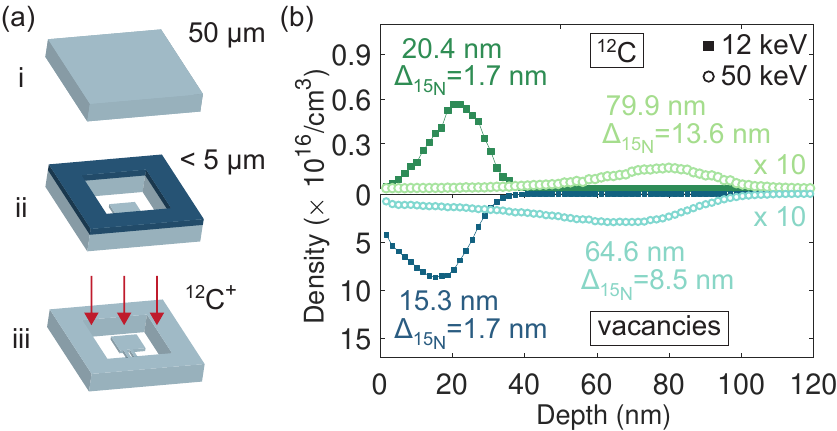}
    \caption{Key aspects of the carbon implantation post-fabrication procedure. (a) Main fabrication steps: (i) A bulk diamond is used as starting material. (ii) Thin microstructures are patterned on the front side of the diamond via lithography and etching. The structures are then released by a back-side etch through a quartz mask. (iii) The microstructures are implanted with $^{12}$C$^+$ from the front side. (b) SRIM-simulated density of implanted $^{12}$C$^+$ ions (top) and resulting vacancies (bottom) for an implantation energy of 12 keV (squares) and 50 keV (circles), as a function of the depth. A fluence of \unit[$1\cdot10^{10}$]{ions/cm$^2$} and \unit[$5\cdot10^8$]{ions/cm$^2$} was taken for the implantation energy of 12 keV and 50 keV, respectively. The depths corresponding to distribution maxima are stated together with the difference $\Delta_{^{15}N}$ in depth compared to implanting $^{15}$N$^+$ with the same energy and fluence. The 50 keV data were multiplied by a factor of 10 for better visibility.}
    \label{fig:fig1}
\end{figure}

On account of its permanent electric dipole moment, the NV's excited state is sensitive to charge noise and thus prone to optical linewidth broadening, especially in microfabricated diamond~\cite{Ruf2019a}. Lifetime-limited optical linewidths, \unit[$\sim$\,13]{MHz}, have therefore been measured only on native NVs in bulk diamond~\cite{Tamarat2006}. It is important to differentiate between two cases of reported linewidths: the ones measured under repeated off-resonant repumping of the charge- and spin state, which we refer to as extrinsically broadened linewidths, and the ones measured between repumping pulses, which we refer to as dephasing linewidths. In the former case, charge noise triggered by the repump pulses induces spectral jumps and results in a broadened linewidth, while in the latter case this contribution is avoided. Recent work demonstrated close to lifetime-limited dephasing linewidths (\unit[$<$\,60]{MHz}) for implanted NVs when using conditional charge state repumping~\cite{Chakravarthi2021}. Under repeated repumping, most implanted centers in bulk instead exhibit extrinsically broadened linewidths on the order of \unit[100]{MHz}, about ten times the lifetime limit~\cite{Chu2014a,Riedel2017c}. For micro- and nano-fabricated diamond, the sensitivity to fluctuating electric fields results in a further deterioration of the optical quality of the NVs: remaining lattice damage from implantation as well as surface and subsurface charge traps introduced during etching conspire such that extrinsically broadened linewidths reaching up to several GHz are observed~\cite{Faraon2012,Li2015,Riedel2017c,Chakravarthi2021,Orphal-Kobin2022}. Furthermore, studies have shown that NVs created from implanted nitrogen have worse optical properties compared to NVs created from diamond-native nitrogen and irradiation-induced vacancies, in that they form a separate, broad-linewidth distribution~\cite{VanDam2019,Kasperczyk2020}. This has led to a number of studies exploring different NV creation methods where nitrogen ion implantation is avoided, including laser writing~\cite{Chen2017b,Yurgens2021} and electron irradiation~\cite{Ruf2019a}. Such approaches have resulted in improved NV optical coherence in bulk and, in the case of electron irradiation, also in \unit[3.8]{\textmu m}-thick diamond membranes, but at the cost of loss of depth control (electron irradiation) and the need for highly specialized setups (laser writing). In general, NV optical linewidths have been shown to increase drastically for structure thicknesses less than \unit[3.8]{\textmu m}, with only a few cases of narrow lines reported in thinner structures~\cite{Kasperczyk2020,Lekavicius2019}.

In this study, we explore a refined method of NV creation, in which we avoid implantation of nitrogen ions and instead use $^{12}$C$^+$ implantation to create vacancies~\cite{Naydenov2010}. The NVs are formed by recombination of the created vacancies with native nitrogen in the diamond lattice. Using carbon as a substitute for nitrogen retains the ability to create NVs with nanoscale depth resolution (unlike electron irradiation or optical writing), without creating a broad-linewidth population (unlike nitrogen implantation). Using $^{12}$C$^+$, which has a mass comparable to $^{15}$N$^+$, for the implantation further ensures that the depth distribution of vacancies is very similar to the one created by implantation of nitrogen. Since the lattice consists mainly of $^{12}$C, this further ensures that no heteroatoms are introduced into the lattice. In contrast to prior work, we opt for implantation post-fabrication (IPF), implanting the ions \emph{after} fabrication of the microstructures, following the approach used by Kasperczyk \emph{et al.}~\cite{Kasperczyk2020}. The rationale behind this reversal is to avoid exposing already formed NVs to the potentially deleterious effects of fabrication, in particular the aggressive dry etching steps and electron beam lithography~\cite{Schwartz2012,Kim2012}. 


\begin{figure}
    \includegraphics{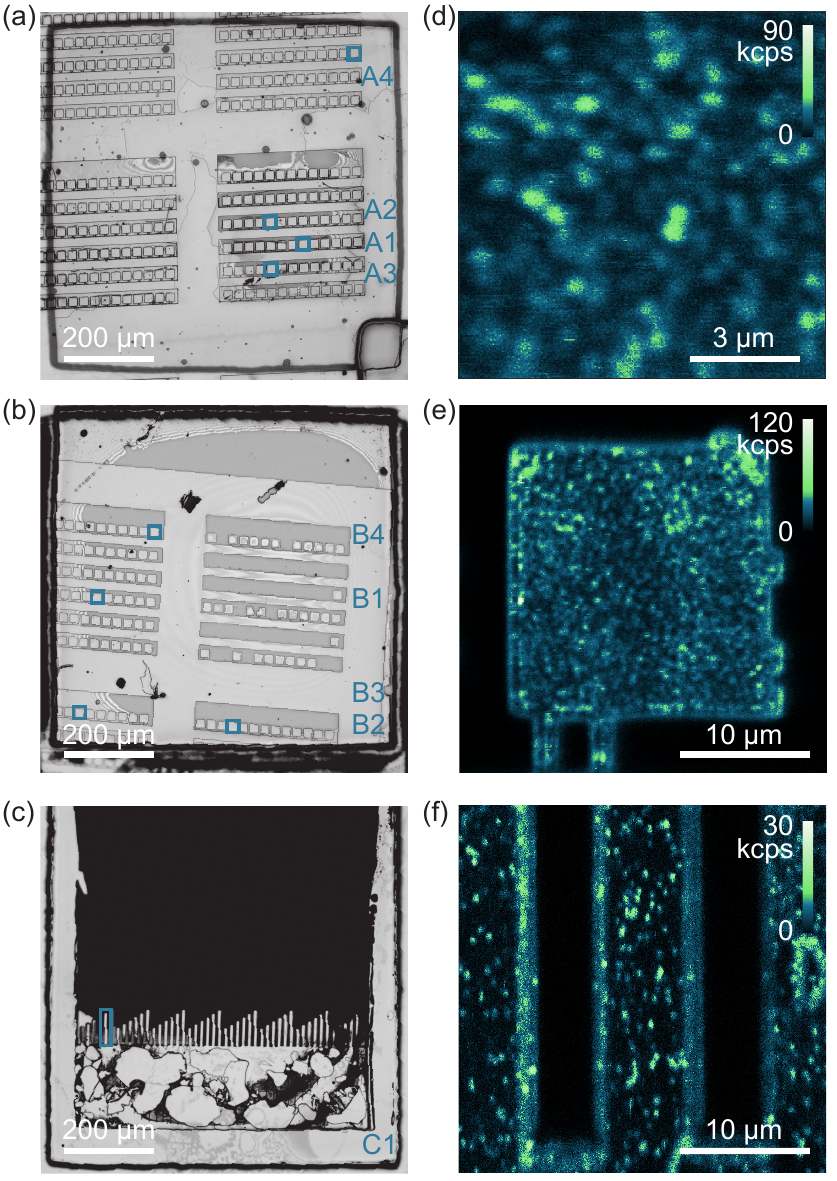}
    \caption{Sample overview. (a)-(c) Brightfield microscope images of samples A-C, respectively, with labels indicating the regions which were sampled for linewidth measurements. The areas shown in the images correspond to the apertures in the quartz masks used in the etching process. (d-f) Representative confocal fluorescence image of NVs in region A2, B1 and C1, respectively.}
    \label{fig:fig2}
\end{figure}

We fabricate our samples from electronic grade (100) single-crystal diamond ([N] $<$\,5 ppb, [B] $<$\,1 ppb, Element Six), and follow a previously developed fabrication procedure~\cite{Appel2016,Challier2018,Riedel2017c}. The main steps of the process flow are illustrated in Fig.\,\ref{fig:fig1}(a). We create \unit[$<$\,5]{\textmu m}-thin microstructures out of \unit[$\sim$\,50]{\textmu m} bulk diamond via electron beam lithography and inductively coupled plasma reactive ion etching (ICP-RIE). The etching is carried out through the aperture (typically \unit[1-by-1]{mm$^2$}) of a quartz mask until the structures within the aperture become free-standing. A combination of the etching process, mask geometry and mask position results in a thickness gradient: the structures in the center of the aperture are up to several micrometers thinner compared to the structures at its edges. The thickness gradient varies in its extent from sample to sample. Once the microstructures have been created, a thin chromium/gold layer is evaporated at a slight angle onto the back surface to reduce charging during implantation. Finally, $^{12}$C$^+$ is implanted into the microstructures from the front side at an angle of 7$^\circ$, and NVs are created after stripping the metal layer and performing a multi-step annealing process similar to the one described in Ref.~\cite{Appel2016}.

As can be seen in Fig.\,\ref{fig:fig1}(b), Stopping and Range of Ions in Matter (SRIM)-simulations for $^{12}$C$^+$ show that the ion- and vacancy distributions after implantation differ only very marginally from implantation with $^{15}$N$^+$. For both \unit[12]{keV} and \unit[50]{keV} implantation energies, the difference in implantation depth of ions (vacancies) inside the diamond, defined as $\Delta_{^{15}N} = |D^{i(v)}_{^{15}N}-D^{i(v)}_{^{12}C}|$ with $D^{i(v)}$ being the depth corresponding to the maximum of the distributions, results in a relative depth difference $\Delta_{^{15}N}/(D^{i(v)}_{^{15}N})$ below 20\% (15\%). We choose the implantation energies \unit[12]{keV} and \unit[50]{keV} for two different purposes: the former to create shallow NVs suitable for optomechanics experiments requiring coherent, near-surface NVs in cantilevers~\cite{Barfuss2015}, and the latter to create deeper NVs located at a vacuum electric field antinode in an open microcavity~\cite{Riedel2017c}. As can be seen in Fig.\,\ref{fig:fig1}(b), we expect the majority of the resulting NVs to be distributed between 15.3 and \unit[20.4]{nm}, and between 64.6 and \unit[79.9]{nm} from the top surface of the diamond, respectively. According to our simulations, the expected vacancy yield per implanted ion is virtually the same for carbon and nitrogen when implanting with \unit[12~(50)]{keV}: \unit[74~(175)]{vacancies/ion} for $^{15}$N$^+$ and \unit[68~(151)]{vacancies/ion} for $^{12}$C$^+$.

Following this procedure, we fabricate three samples, denoted sample A, B, and C. Samples A and B have square, \unit[20-by-20]{\textmu m$^2$} microstructures and sample C has rectangular, \unit[5-10]{\textmu m} wide and \unit[50-100]{\textmu m} long microstructures. We create NVs at different depths via carbon IPF and study the linewidths of the resulting NVs as a function of the microstructure thickness. Sample A and B are both $^{12}$C$^+$-implanted with an energy of \unit[55]{keV} and a fluence of \unit[$5\cdot10^8$]{ions/cm$^2$}, while sample C is implanted with an energy of \unit[12]{keV} and a fluence of \unit[$1\cdot10^{10}$]{ions/cm$^2$}. 

We image the emitters using a home-built confocal microscope, where we use a \unit[532]{nm} diode-pumped solid-state laser for samples A and B and a \unit[515]{nm} diode laser module for sample C for off-resonant excitation while detecting the resulting photoluminescence (PL) after filtering by a long-pass filter (Semrock, 594 nm RazorEdge) on a single-photon counting photodetector (Excelitas). Figure \ref{fig:fig2} shows brightfield images as well as confocal scans of representative regions on the three samples. 

\begin{figure}[t!]
    \includegraphics{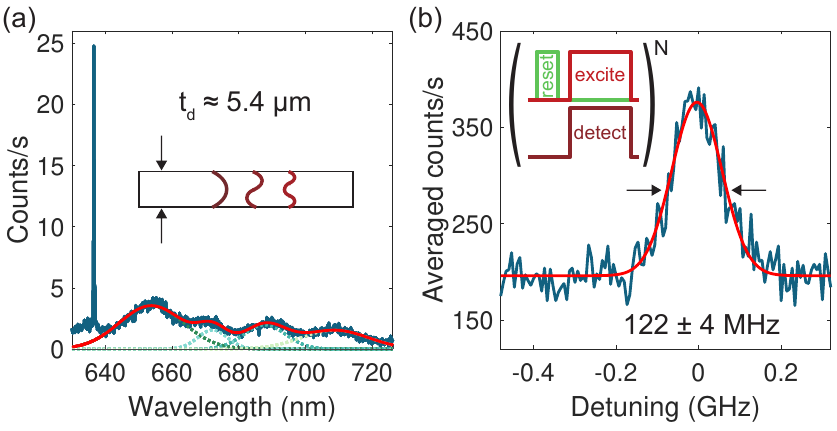}
    \caption{Measurement examples. (a) PL of an NV in sample A. The PSB has been fitted to extract the wavelengths that give constructive interference within the microstructure (dotted lines), in turn providing the structure thickness (\unit[5.4]{\textmu m}). The inset illustrates the etaloning effect allowing for thickness determination. (b) PLE of the same NV as in (a). The inset shows the PLE measurement sequence.}
    \label{fig:fig3}
\end{figure}

We record NV photoluminescence (PL) spectra on a cryogenically cooled CCD camera coupled to a grating spectrometer (Princeton Instruments). An example spectrum is shown in Fig.\,\ref{fig:fig3}(a). The thickness of the measured microstructures is determined either from the spectra by fitting the amplitude modulation of the NV's phonon sideband (PSB) arising from interference in the diamond, or by direct measurement with a commercial laser confocal microscope (Keyence VK-X1100). We choose measurement sites on the three samples covering a range of different thicknesses. The chosen regions are indicated in Fig.\,\ref{fig:fig2}(a)-(c): in sample A, thicknesses span \unit[1.9-4.6]{\textmu m}, and in sample B \unit[1.9-4.9]{\textmu m}. The free-standing microstructures in sample C are \unit[2.5]{\textmu m} thick and we also measure in a neighboring bulk region with a thickness of approximately \unit[50]{\textmu m} (not indicated).

We characterize the NV ZPL optical linewidths through photoluminescence excitation (PLE) measurements in either a liquid helium bath (CryoVac) or a closed-loop cold-finger (attocube attoDRY800) cryostat. We resonantly excite the NVs by sweeping a tunable narrow-linewidth external cavity diode laser (New Focus Velocity TLB-6704) locked to a wavemeter (HighFinesse WS-U) across the ZPL and collect the photons emitted into the PSB with a high-numerical aperture objective (Partec 50x, 0.82 NA, or Olympus LMPLFLN 100x, 0.8 NA). We minimize power broadening in the resulting PLE spectra by measuring at the lowest possible resonant laser power that still gives an adequate signal-to-noise ratio. Typical resonant laser powers are on the order of \unit[100]{nW}, which leads to a power broadening of up to three times the homogeneous linewidth yet remains marginal compared to our measured linewidths~\cite{Yurgens2021}. The measurement sequence consists of a \unit[1-2]{\textmu s} green initialization pulse to reset the charge- and spin state, followed by a resonant pulse (\unit[637]{nm} $\pm$ detuning, \unit[5-10]{\textmu s}) during which PSB counts are accumulated. The sequence is repeated at a \unit[100]{kHz} rate for a duration of \unit[10]{ms} for each laser detuning. An example PLE measurement from the average of 100 scans is shown in Fig.\,\ref{fig:fig3}(b). The residual PLE background arises via leakage of photons generated during the green repumping pulse into the counting window. We note that by systematically using an off-resonant green initialization pulse, our experiment yields a measurement of the extrinsically broadened linewidths, i.e., linewidths including spectral jumps caused by the repump pulses. 

\begin{figure}
    \includegraphics{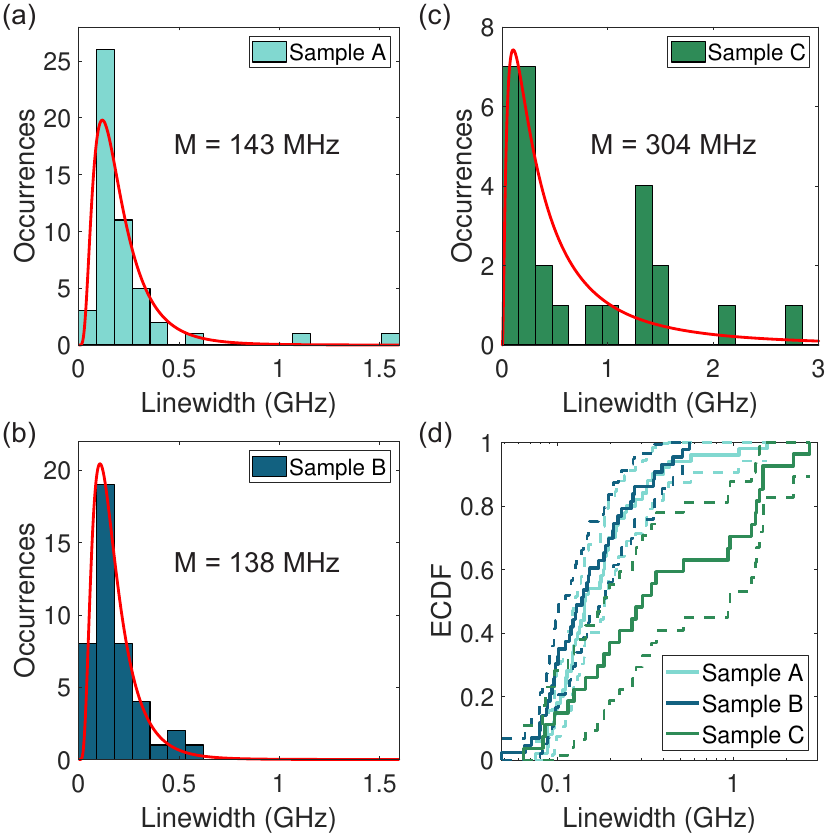}
    \caption{Linewidth statistics. (a)-(c) Histograms over the measured linewidths for sample A (implantation energy 50 keV), B (implantation energy 50 keV) and C (implantation energy 12 keV), respectively. Solid red lines represent log-normal fits and the median M is stated for each distribution. (d) Empirical cumulative distribution functions for all samples, plotted on a logarithmic scale for better data visibility. The dashed lines represent a 95\% confidence interval.} \vspace{-2mm}
    \label{fig:fig4}
\end{figure}

Figure \ref{fig:fig4} summarizes the measured PLE linewidths in samples A, B, and C. We measure a median (mean) Gaussian-fitted full width at half maximum (FWHM) linewidth of \unit[143]{MHz} (\unit[227]{MHz}), \unit[138]{MHz} (\unit[181]{MHz}) and \unit[304]{MHz} (\unit[691]{MHz}) for the three samples, respectively. The data are well-described by lognormal distributions~\cite{Kasperczyk2020}. We argue that the median is a better figure-of-merit than the mean as the linewidth distributions are asymmetric. A few broad-linewidth NVs in the tail of the distribution will increase the mean but barely affect the amount of preselection needed in any application requiring narrow-linewidth NVs. In this case, the median is more informative, providing the linewidth value for which half of the population is equal or narrower. Figure \ref{fig:fig4}(d) shows the corresponding empirical cumulative distribution functions, indicating the fraction of measured linewidths below or equal to a certain value. 

In order to provide a quantitative threshold under which NV linewidths can be described as narrow, we base ourselves on the projected indistinguishability between two ZPL photons -- a cornerstone for remote entanglement generation~\cite{Bernien2013}. Two-photon interference with high (0.9) visibility can still be obtained despite broadening at the cost of temporal filtering, down to the typical \unit[$\sim$\,300]{ps} timing resolution of silicon single photon counting diodes~\cite{Legero2003, Legero2006,Kambs2018}. In our case, we cannot discriminate between broadening due to spectral jumps and broadening due to pure dephasing. Considering the latter as the sole contribution leads to the worst-case scenario, in which temporal selection is technically possible for measured linewidths \unit[$\lesssim$\,150]{MHz}. In samples A and B, which we group as the two samples with the same implantation parameters, 54\% of the measured linewidths are below \unit[150]{MHz}. In sample C, only 26\% of the linewidths are below \unit[150]{MHz}, and the median linewidth is about two times higher. This difference can be explained by the shallower implantation depth of sample C and a correspondingly larger contribution of surface-related charge noise to the linewidth broadening~\cite{Chakravarthi2021}. For the three samples combined, including bulk, we measure 48\% of linewidths below \unit[150]{MHz}. 

\begin{figure}
    \includegraphics{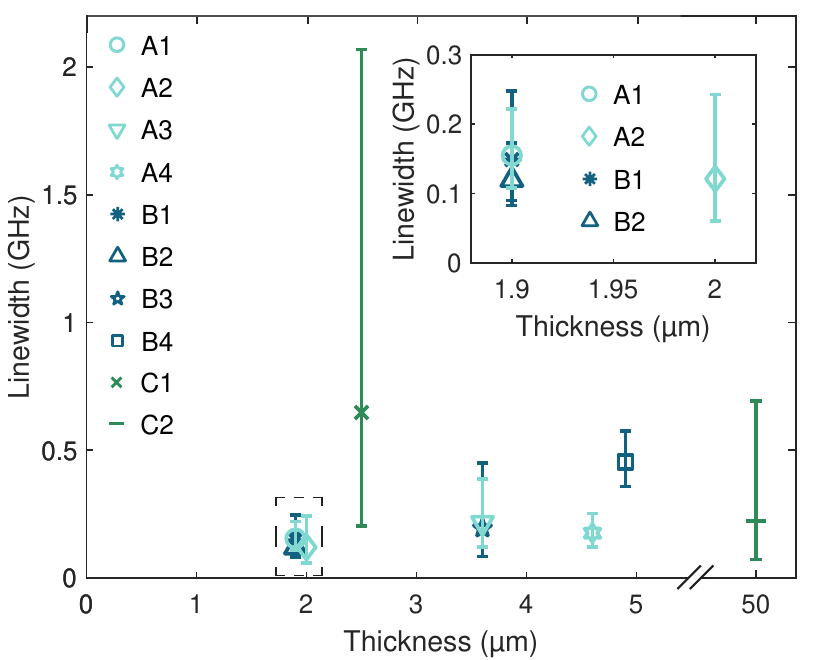}
    \caption{Median linewidth as a function of diamond thickness. The error bars correspond to the geometrical standard deviation of a log-normal fit to each distribution. The data points are labelled according to the sample and corresponding microstructure, in order of increasing thickness within each sample. The inset shows a close-up of the median linewidths of the four thinnest microstructures.} \vspace{-2mm}
    \label{fig:fig5}
\end{figure}

Figure \ref{fig:fig5} demonstrates the statistical changes in linewidths as a function of the thickness of the measured structure. We see no general trend of increasing mean (not shown) or median linewidth with decreasing structure thickness, in contrast to previous studies~\cite{Ruf2019a,Kasperczyk2020,Riedel2017c}. Sample C shows broader lines than the other two samples regardless of the sampled thickness, which can again be attributed to the shallower NV depth in this sample. Considering all the sampled microstructures, spanning \unit[1.9-4.9]{\textmu m} in thickness, 52\% of the NVs have a linewidth below \unit[150]{MHz}. No thinner regions were created on samples A and C. On sample B, the fraction of NVs exhibiting optically addressable ZPLs dropped significantly for structures less than \unit[1.9]{\textmu m} thick, which precluded further studies. We note that narrow, \unit[$<$\,250]{MHz}, extrinsically broadened linewidths have previously not been measured consistently in diamond microstructures thinner than \unit[3.8]{\textmu m}~\cite{Ruf2019a,Lekavicius2019,Kasperczyk2020}. IPF with nitrogen as employed by Kasperczyk \emph{et al.} described a few cases of narrow linewidths in thin structures; as we show here, going one step further and substituting nitrogen for carbon avoids the formation of broader-linewidth NVs from implanted nitrogen and therefore results in only the native-nitrogen, narrow-linewidth NV distribution.

We speculate that a possible explanation for the positive effect of IPF on NV optical linewidths is that the dynamics of NV formation are affected by preexisting lattice damage. Performing etching and electron beam lithography on structures with existing emitters imposes lattice damage on the complete structure, while implantation after the microfabrication might preferentially create emitters on sites with less lattice disorder. Dedicated studies are however required to single out the effect of IPF from carbon substitution alone.  

In conclusion, we demonstrate that carbon IPF creates NVs with on average lower charge-noise and reduced spectral diffusion compared to other NV creation approaches. We show that substituting nitrogen with carbon during implantation results in similar NV densities for moderate implantation fluences that allow optical isolation of single emitters. We measure narrow linewidths with high probability: 48\% of the measured NVs have an extrinsically broadened linewidth below \unit[150]{MHz}, and this applies across the full measured range of diamond microstructure thicknesses down to \unit[1.9]{\textmu m}. We show that by implanting an ion species other than nitrogen, the creation of NVs displaying extreme linewidth broadening is avoided, confirming earlier conjectures~\cite{VanDam2019, Kasperczyk2020}. Moreover, we demonstrate that our method is easy to implement, is reproducible and provides an advantage over conventional fabrication methods for different samples and implantation parameters.  

Our method should prove useful for any application relying on optically coherent NVs, both in bulk and in microstructured diamond. As an example, coupling the emission of such NVs (\unit[$\sim$\,150]{MHz} linewidth, \unit[$\sim$\,2]{\textmu m} thick microstructure) to an open microcavity with a finesse of a few thousand would increase the fraction of photons emitted into the ZPL by more than an order of magnitude~\cite{Riedel2017c}. The success probability of spin-spin entanglement between remote NVs would therefore increase by over two orders of magnitude compared to the state-of-the-art when using the Barrett-Kok protocol~\cite{Ruf2019a,Hensen2015,Barrett2005}. This would dramatically improve the generation rate of entangled qubit pairs and remove an important roadblock toward quantum networks beyond few-node prototypes.

\begin{acknowledgments}
We thank Tomasz Jakubczyk for fruitful discussions. We acknowledge financial support from the National Centre of Competence in Research (NCCR) Quantum Science and Technology (QSIT), a competence center funded by the Swiss National Science Foundation (SNF), and by SNF project No.\,188521. J.A.Z. acknowledges financial support from the Swiss Nanoscience Institute (SNI). A.C. acknowledges financial support from the Quantum Science and Technologies at the European Campus (QUSTEC) project of the European Union’s Horizon 2020 research and innovation programme under the Marie Skłodowska-Curie grant agreement No.\,847471.
\end{acknowledgments}

\bibliography{Paper_CarbonIPF_arXiv} 

\begin{thebibliography}{46}%
\makeatletter
\providecommand \@ifxundefined [1]{%
 \@ifx{#1\undefined}
}%
\providecommand \@ifnum [1]{%
 \ifnum #1\expandafter \@firstoftwo
 \else \expandafter \@secondoftwo
 \fi
}%
\providecommand \@ifx [1]{%
 \ifx #1\expandafter \@firstoftwo
 \else \expandafter \@secondoftwo
 \fi
}%
\providecommand \natexlab [1]{#1}%
\providecommand \enquote  [1]{``#1''}%
\providecommand \bibnamefont  [1]{#1}%
\providecommand \bibfnamefont [1]{#1}%
\providecommand \citenamefont [1]{#1}%
\providecommand \href@noop [0]{\@secondoftwo}%
\providecommand \href [0]{\begingroup \@sanitize@url \@href}%
\providecommand \@href[1]{\@@startlink{#1}\@@href}%
\providecommand \@@href[1]{\endgroup#1\@@endlink}%
\providecommand \@sanitize@url [0]{\catcode `\\12\catcode `\$12\catcode
  `\&12\catcode `\#12\catcode `\^12\catcode `\_12\catcode `\%12\relax}%
\providecommand \@@startlink[1]{}%
\providecommand \@@endlink[0]{}%
\providecommand \url  [0]{\begingroup\@sanitize@url \@url }%
\providecommand \@url [1]{\endgroup\@href {#1}{\urlprefix }}%
\providecommand \urlprefix  [0]{URL }%
\providecommand \Eprint [0]{\href }%
\providecommand \doibase [0]{https://doi.org/}%
\providecommand \selectlanguage [0]{\@gobble}%
\providecommand \bibinfo  [0]{\@secondoftwo}%
\providecommand \bibfield  [0]{\@secondoftwo}%
\providecommand \translation [1]{[#1]}%
\providecommand \BibitemOpen [0]{}%
\providecommand \bibitemStop [0]{}%
\providecommand \bibitemNoStop [0]{.\EOS\space}%
\providecommand \EOS [0]{\spacefactor3000\relax}%
\providecommand \BibitemShut  [1]{\csname bibitem#1\endcsname}%
\let\auto@bib@innerbib\@empty
\bibitem [{\citenamefont {Awschalom}\ \emph {et~al.}(2018)\citenamefont
  {Awschalom}, \citenamefont {Hanson}, \citenamefont {Wrachtrup},\ and\
  \citenamefont {Zhou}}]{Awschalom2018}%
  \BibitemOpen
  \bibfield  {author} {\bibinfo {author} {\bibfnamefont {D.~D.}\ \bibnamefont
  {Awschalom}}, \bibinfo {author} {\bibfnamefont {R.}~\bibnamefont {Hanson}},
  \bibinfo {author} {\bibfnamefont {J.}~\bibnamefont {Wrachtrup}},\ and\
  \bibinfo {author} {\bibfnamefont {B.~B.}\ \bibnamefont {Zhou}},\ }\bibfield
  {title} {\bibinfo {title} {Quantum technologies with optically interfaced
  solid-state spins},\ }\href@noop {} {\bibfield  {journal} {\bibinfo
  {journal} {Nat. Photonics}\ }\textbf {\bibinfo {volume} {12}},\ \bibinfo
  {pages} {516} (\bibinfo {year} {2018})}\BibitemShut {NoStop}%
\bibitem [{\citenamefont {Degen}\ \emph {et~al.}(2017)\citenamefont {Degen},
  \citenamefont {Reinhard},\ and\ \citenamefont {Cappellaro}}]{Degen2017}%
  \BibitemOpen
  \bibfield  {author} {\bibinfo {author} {\bibfnamefont {C.~L.}\ \bibnamefont
  {Degen}}, \bibinfo {author} {\bibfnamefont {F.}~\bibnamefont {Reinhard}},\
  and\ \bibinfo {author} {\bibfnamefont {P.}~\bibnamefont {Cappellaro}},\
  }\bibfield  {title} {\bibinfo {title} {Quantum sensing},\ }\href@noop {}
  {\bibfield  {journal} {\bibinfo  {journal} {Rev. Mod. Phys.}\ }\textbf
  {\bibinfo {volume} {89}},\ \bibinfo {pages} {035002} (\bibinfo {year}
  {2017})}\BibitemShut {NoStop}%
\bibitem [{\citenamefont {Balasubramanian}\ \emph {et~al.}(2009)\citenamefont
  {Balasubramanian}, \citenamefont {Neumann}, \citenamefont {Twitchen},
  \citenamefont {Markham}, \citenamefont {Kolesov}, \citenamefont {Mizuochi},
  \citenamefont {Isoya}, \citenamefont {Achard}, \citenamefont {Beck},
  \citenamefont {Tissler}, \citenamefont {Jacques}, \citenamefont {Hemmer},
  \citenamefont {Jelezko},\ and\ \citenamefont
  {Wrachtrup}}]{Balasubramanian2009}%
  \BibitemOpen
  \bibfield  {author} {\bibinfo {author} {\bibfnamefont {G.}~\bibnamefont
  {Balasubramanian}}, \bibinfo {author} {\bibfnamefont {P.}~\bibnamefont
  {Neumann}}, \bibinfo {author} {\bibfnamefont {D.}~\bibnamefont {Twitchen}},
  \bibinfo {author} {\bibfnamefont {M.}~\bibnamefont {Markham}}, \bibinfo
  {author} {\bibfnamefont {R.}~\bibnamefont {Kolesov}}, \bibinfo {author}
  {\bibfnamefont {N.}~\bibnamefont {Mizuochi}}, \bibinfo {author}
  {\bibfnamefont {J.}~\bibnamefont {Isoya}}, \bibinfo {author} {\bibfnamefont
  {J.}~\bibnamefont {Achard}}, \bibinfo {author} {\bibfnamefont
  {J.}~\bibnamefont {Beck}}, \bibinfo {author} {\bibfnamefont {J.}~\bibnamefont
  {Tissler}}, \bibinfo {author} {\bibfnamefont {V.}~\bibnamefont {Jacques}},
  \bibinfo {author} {\bibfnamefont {P.~R.}\ \bibnamefont {Hemmer}}, \bibinfo
  {author} {\bibfnamefont {F.}~\bibnamefont {Jelezko}},\ and\ \bibinfo {author}
  {\bibfnamefont {J.}~\bibnamefont {Wrachtrup}},\ }\bibfield  {title} {\bibinfo
  {title} {Ultralong spin coherence time in isotopically engineered diamond},\
  }\href@noop {} {\bibfield  {journal} {\bibinfo  {journal} {Nature Mat.}\
  }\textbf {\bibinfo {volume} {8}},\ \bibinfo {pages} {383} (\bibinfo {year}
  {2009})}\BibitemShut {NoStop}%
\bibitem [{\citenamefont {Bar-Gill}\ \emph {et~al.}(2013)\citenamefont
  {Bar-Gill}, \citenamefont {Pham}, \citenamefont {Jarmola}, \citenamefont
  {Budker},\ and\ \citenamefont {Walsworth}}]{Bar-Gill2013}%
  \BibitemOpen
  \bibfield  {author} {\bibinfo {author} {\bibfnamefont {N.}~\bibnamefont
  {Bar-Gill}}, \bibinfo {author} {\bibfnamefont {L.~M.}\ \bibnamefont {Pham}},
  \bibinfo {author} {\bibfnamefont {A.}~\bibnamefont {Jarmola}}, \bibinfo
  {author} {\bibfnamefont {D.}~\bibnamefont {Budker}},\ and\ \bibinfo {author}
  {\bibfnamefont {R.~L.}\ \bibnamefont {Walsworth}},\ }\bibfield  {title}
  {\bibinfo {title} {Solid-state electronic spin coherence time approaching one
  second},\ }\href@noop {} {\bibfield  {journal} {\bibinfo  {journal} {Nat.
  Commun.}\ }\textbf {\bibinfo {volume} {4}},\ \bibinfo {pages} {1743}
  (\bibinfo {year} {2013})}\BibitemShut {NoStop}%
\bibitem [{\citenamefont {Abobeih}\ \emph {et~al.}(2018)\citenamefont
  {Abobeih}, \citenamefont {Cramer}, \citenamefont {Bakker}, \citenamefont
  {Kalb}, \citenamefont {Markham}, \citenamefont {Twitchen},\ and\
  \citenamefont {Taminiau}}]{Abobeih2018}%
  \BibitemOpen
  \bibfield  {author} {\bibinfo {author} {\bibfnamefont {M.~H.}\ \bibnamefont
  {Abobeih}}, \bibinfo {author} {\bibfnamefont {J.}~\bibnamefont {Cramer}},
  \bibinfo {author} {\bibfnamefont {M.~A.}\ \bibnamefont {Bakker}}, \bibinfo
  {author} {\bibfnamefont {N.}~\bibnamefont {Kalb}}, \bibinfo {author}
  {\bibfnamefont {M.}~\bibnamefont {Markham}}, \bibinfo {author} {\bibfnamefont
  {D.~J.}\ \bibnamefont {Twitchen}},\ and\ \bibinfo {author} {\bibfnamefont
  {T.~H.}\ \bibnamefont {Taminiau}},\ }\bibfield  {title} {\bibinfo {title}
  {One-second coherence for a single electron spin coupled to a multi-qubit
  nuclear-spin environment},\ }\href@noop {} {\bibfield  {journal} {\bibinfo
  {journal} {Nat. Commun.}\ }\textbf {\bibinfo {volume} {9}},\ \bibinfo {pages}
  {1} (\bibinfo {year} {2018})}\BibitemShut {NoStop}%
\bibitem [{\citenamefont {Bradley}\ \emph {et~al.}(2019)\citenamefont
  {Bradley}, \citenamefont {Randall}, \citenamefont {Abobeih}, \citenamefont
  {Berrevoets}, \citenamefont {Degen}, \citenamefont {Bakker}, \citenamefont
  {Markham}, \citenamefont {Twitchen},\ and\ \citenamefont
  {Taminiau}}]{Bradley2019}%
  \BibitemOpen
  \bibfield  {author} {\bibinfo {author} {\bibfnamefont {C.~E.}\ \bibnamefont
  {Bradley}}, \bibinfo {author} {\bibfnamefont {J.}~\bibnamefont {Randall}},
  \bibinfo {author} {\bibfnamefont {M.~H.}\ \bibnamefont {Abobeih}}, \bibinfo
  {author} {\bibfnamefont {R.~C.}\ \bibnamefont {Berrevoets}}, \bibinfo
  {author} {\bibfnamefont {M.~J.}\ \bibnamefont {Degen}}, \bibinfo {author}
  {\bibfnamefont {M.~A.}\ \bibnamefont {Bakker}}, \bibinfo {author}
  {\bibfnamefont {M.}~\bibnamefont {Markham}}, \bibinfo {author} {\bibfnamefont
  {D.~J.}\ \bibnamefont {Twitchen}},\ and\ \bibinfo {author} {\bibfnamefont
  {T.~H.}\ \bibnamefont {Taminiau}},\ }\bibfield  {title} {\bibinfo {title} {A
  ten-qubit solid-state spin register with quantum memory up to one minute},\
  }\href@noop {} {\bibfield  {journal} {\bibinfo  {journal} {Phys. Rev. X}\
  }\textbf {\bibinfo {volume} {9}},\ \bibinfo {pages} {031045} (\bibinfo {year}
  {2019})}\BibitemShut {NoStop}%
\bibitem [{\citenamefont {Doherty}\ \emph {et~al.}(2013)\citenamefont
  {Doherty}, \citenamefont {Manson}, \citenamefont {Delaney}, \citenamefont
  {Jelezko}, \citenamefont {Wrachtrup},\ and\ \citenamefont
  {Hollenberg}}]{Doherty2013}%
  \BibitemOpen
  \bibfield  {author} {\bibinfo {author} {\bibfnamefont {M.~W.}\ \bibnamefont
  {Doherty}}, \bibinfo {author} {\bibfnamefont {N.~B.}\ \bibnamefont {Manson}},
  \bibinfo {author} {\bibfnamefont {P.}~\bibnamefont {Delaney}}, \bibinfo
  {author} {\bibfnamefont {F.}~\bibnamefont {Jelezko}}, \bibinfo {author}
  {\bibfnamefont {J.}~\bibnamefont {Wrachtrup}},\ and\ \bibinfo {author}
  {\bibfnamefont {L.~C.~L.}\ \bibnamefont {Hollenberg}},\ }\bibfield  {title}
  {\bibinfo {title} {The nitrogen-vacancy colour centre in diamond},\
  }\href@noop {} {\bibfield  {journal} {\bibinfo  {journal} {Phys. Rep.}\
  }\textbf {\bibinfo {volume} {528}},\ \bibinfo {pages} {1} (\bibinfo {year}
  {2013})}\BibitemShut {NoStop}%
\bibitem [{\citenamefont {Thiel}\ \emph {et~al.}(2019)\citenamefont {Thiel},
  \citenamefont {Wang}, \citenamefont {Tschudin}, \citenamefont {Rohner},
  \citenamefont {Guti\'{e}rrez-Lexama}, \citenamefont {Ubrig}, \citenamefont
  {Gibertini}, \citenamefont {Giannini}, \citenamefont {Morpurgo},\ and\
  \citenamefont {Maletinsky}}]{Thiel2019}%
  \BibitemOpen
  \bibfield  {author} {\bibinfo {author} {\bibfnamefont {L.}~\bibnamefont
  {Thiel}}, \bibinfo {author} {\bibfnamefont {Z.}~\bibnamefont {Wang}},
  \bibinfo {author} {\bibfnamefont {M.~A.}\ \bibnamefont {Tschudin}}, \bibinfo
  {author} {\bibfnamefont {D.}~\bibnamefont {Rohner}}, \bibinfo {author}
  {\bibfnamefont {I.}~\bibnamefont {Guti\'{e}rrez-Lexama}}, \bibinfo {author}
  {\bibfnamefont {N.}~\bibnamefont {Ubrig}}, \bibinfo {author} {\bibfnamefont
  {M.}~\bibnamefont {Gibertini}}, \bibinfo {author} {\bibfnamefont
  {E.}~\bibnamefont {Giannini}}, \bibinfo {author} {\bibfnamefont {A.~F.}\
  \bibnamefont {Morpurgo}},\ and\ \bibinfo {author} {\bibfnamefont
  {P.}~\bibnamefont {Maletinsky}},\ }\bibfield  {title} {\bibinfo {title}
  {Probing magnetism in 2d materials at the nanoscale with single-spin
  microscopy},\ }\href@noop {} {\bibfield  {journal} {\bibinfo  {journal}
  {Science}\ }\textbf {\bibinfo {volume} {364}},\ \bibinfo {pages} {973}
  (\bibinfo {year} {2019})}\BibitemShut {NoStop}%
\bibitem [{\citenamefont {Hedrich}\ \emph {et~al.}(2021)\citenamefont
  {Hedrich}, \citenamefont {Wagner}, \citenamefont {Pylypovskyi}, \citenamefont
  {Shields}, \citenamefont {Kosub}, \citenamefont {Sheka}, \citenamefont
  {Makarov},\ and\ \citenamefont {Maletinsky}}]{Hedrich2021}%
  \BibitemOpen
  \bibfield  {author} {\bibinfo {author} {\bibfnamefont {N.}~\bibnamefont
  {Hedrich}}, \bibinfo {author} {\bibfnamefont {K.}~\bibnamefont {Wagner}},
  \bibinfo {author} {\bibfnamefont {O.~V.}\ \bibnamefont {Pylypovskyi}},
  \bibinfo {author} {\bibfnamefont {B.~J.}\ \bibnamefont {Shields}}, \bibinfo
  {author} {\bibfnamefont {T.}~\bibnamefont {Kosub}}, \bibinfo {author}
  {\bibfnamefont {D.~D.}\ \bibnamefont {Sheka}}, \bibinfo {author}
  {\bibfnamefont {D.}~\bibnamefont {Makarov}},\ and\ \bibinfo {author}
  {\bibfnamefont {P.}~\bibnamefont {Maletinsky}},\ }\bibfield  {title}
  {\bibinfo {title} {Nanoscale mechanics of antiferromagnetic domain walls},\
  }\href@noop {} {\bibfield  {journal} {\bibinfo  {journal} {Nat. Phys.}\
  }\textbf {\bibinfo {volume} {17}},\ \bibinfo {pages} {574} (\bibinfo {year}
  {2021})}\BibitemShut {NoStop}%
\bibitem [{\citenamefont {Palm}\ \emph {et~al.}(2022)\citenamefont {Palm},
  \citenamefont {Huxter}, \citenamefont {Welter}, \citenamefont {Ernst},
  \citenamefont {Scheidegger}, \citenamefont {Diesch}, \citenamefont {Chang},
  \citenamefont {Rickhaus}, \citenamefont {Taniguchi}, \citenamefont
  {Watanabe}, \citenamefont {Ensslin},\ and\ \citenamefont {Degen}}]{Palm2022}%
  \BibitemOpen
  \bibfield  {author} {\bibinfo {author} {\bibfnamefont {M.~L.}\ \bibnamefont
  {Palm}}, \bibinfo {author} {\bibfnamefont {W.~S.}\ \bibnamefont {Huxter}},
  \bibinfo {author} {\bibfnamefont {P.}~\bibnamefont {Welter}}, \bibinfo
  {author} {\bibfnamefont {S.}~\bibnamefont {Ernst}}, \bibinfo {author}
  {\bibfnamefont {P.~J.}\ \bibnamefont {Scheidegger}}, \bibinfo {author}
  {\bibfnamefont {S.}~\bibnamefont {Diesch}}, \bibinfo {author} {\bibfnamefont
  {K.}~\bibnamefont {Chang}}, \bibinfo {author} {\bibfnamefont
  {P.}~\bibnamefont {Rickhaus}}, \bibinfo {author} {\bibfnamefont
  {T.}~\bibnamefont {Taniguchi}}, \bibinfo {author} {\bibfnamefont
  {K.}~\bibnamefont {Watanabe}}, \bibinfo {author} {\bibfnamefont
  {K.}~\bibnamefont {Ensslin}},\ and\ \bibinfo {author} {\bibfnamefont {C.~L.}\
  \bibnamefont {Degen}},\ }\bibfield  {title} {\bibinfo {title} {Imaging of
  submicroampere currents in bilayer graphene using a scanning diamond
  magnetometer},\ }\href@noop {} {\bibfield  {journal} {\bibinfo  {journal}
  {Phys. Rev. Applied}\ }\textbf {\bibinfo {volume} {17}},\ \bibinfo {pages}
  {054008} (\bibinfo {year} {2022})}\BibitemShut {NoStop}%
\bibitem [{\citenamefont {Hensen}\ \emph {et~al.}(2015)\citenamefont {Hensen},
  \citenamefont {Bernien}, \citenamefont {Dr{\'{e}au}}, \citenamefont
  {Reiserer}, \citenamefont {Kalb}, \citenamefont {Blok}, \citenamefont
  {Ruitenberg}, \citenamefont {Vermeulen}, \citenamefont {Schouten},
  \citenamefont {Abell{\'{a}}n}, \citenamefont {Amaya}, \citenamefont
  {Pruneri}, \citenamefont {Mitchell}, \citenamefont {Markham}, \citenamefont
  {Twitchen}, \citenamefont {Elkouss}, \citenamefont {Wehner}, \citenamefont
  {Taminiau},\ and\ \citenamefont {Hanson}}]{Hensen2015}%
  \BibitemOpen
  \bibfield  {author} {\bibinfo {author} {\bibfnamefont {B.}~\bibnamefont
  {Hensen}}, \bibinfo {author} {\bibfnamefont {H.}~\bibnamefont {Bernien}},
  \bibinfo {author} {\bibfnamefont {A.~E.}\ \bibnamefont {Dr{\'{e}au}}},
  \bibinfo {author} {\bibfnamefont {A.}~\bibnamefont {Reiserer}}, \bibinfo
  {author} {\bibfnamefont {N.}~\bibnamefont {Kalb}}, \bibinfo {author}
  {\bibfnamefont {M.~S.}\ \bibnamefont {Blok}}, \bibinfo {author}
  {\bibfnamefont {J.}~\bibnamefont {Ruitenberg}}, \bibinfo {author}
  {\bibfnamefont {R.~F.~L.}\ \bibnamefont {Vermeulen}}, \bibinfo {author}
  {\bibfnamefont {R.~N.}\ \bibnamefont {Schouten}}, \bibinfo {author}
  {\bibfnamefont {C.}~\bibnamefont {Abell{\'{a}}n}}, \bibinfo {author}
  {\bibfnamefont {W.}~\bibnamefont {Amaya}}, \bibinfo {author} {\bibfnamefont
  {V.}~\bibnamefont {Pruneri}}, \bibinfo {author} {\bibfnamefont {M.~W.}\
  \bibnamefont {Mitchell}}, \bibinfo {author} {\bibfnamefont {M.}~\bibnamefont
  {Markham}}, \bibinfo {author} {\bibfnamefont {D.~J.}\ \bibnamefont
  {Twitchen}}, \bibinfo {author} {\bibfnamefont {D.}~\bibnamefont {Elkouss}},
  \bibinfo {author} {\bibfnamefont {S.}~\bibnamefont {Wehner}}, \bibinfo
  {author} {\bibfnamefont {T.~H.}\ \bibnamefont {Taminiau}},\ and\ \bibinfo
  {author} {\bibfnamefont {R.}~\bibnamefont {Hanson}},\ }\bibfield  {title}
  {\bibinfo {title} {Loophole-free bell inequality violation using electron
  spins separated by 1.3 kilometres},\ }\href@noop {} {\bibfield  {journal}
  {\bibinfo  {journal} {Nature}\ }\textbf {\bibinfo {volume} {526}},\ \bibinfo
  {pages} {682–686} (\bibinfo {year} {2015})}\BibitemShut {NoStop}%
\bibitem [{\citenamefont {Pompili}\ \emph {et~al.}(2021)\citenamefont
  {Pompili}, \citenamefont {Hermans}, \citenamefont {Baier}, \citenamefont
  {Beukers}, \citenamefont {Humphreys}, \citenamefont {Schouten}, \citenamefont
  {Vermeulen}, \citenamefont {Tiggelman}, \citenamefont {Dos Santos~Martins},\
  and\ \citenamefont {Hanson}}]{Pompili2021}%
  \BibitemOpen
  \bibfield  {author} {\bibinfo {author} {\bibfnamefont {M.}~\bibnamefont
  {Pompili}}, \bibinfo {author} {\bibfnamefont {S.~L.~N.}\ \bibnamefont
  {Hermans}}, \bibinfo {author} {\bibfnamefont {S.}~\bibnamefont {Baier}},
  \bibinfo {author} {\bibfnamefont {H.~K.~C.}\ \bibnamefont {Beukers}},
  \bibinfo {author} {\bibfnamefont {P.~C.}\ \bibnamefont {Humphreys}}, \bibinfo
  {author} {\bibfnamefont {R.~N.}\ \bibnamefont {Schouten}}, \bibinfo {author}
  {\bibfnamefont {R.~F.~L.}\ \bibnamefont {Vermeulen}}, \bibinfo {author}
  {\bibfnamefont {M.~J.}\ \bibnamefont {Tiggelman}}, \bibinfo {author}
  {\bibfnamefont {L.}~\bibnamefont {Dos Santos~Martins}},\ and\ \bibinfo
  {author} {\bibfnamefont {R.}~\bibnamefont {Hanson}},\ }\bibfield  {title}
  {\bibinfo {title} {Realization of a multinode quantum network of remote
  solid-state qubits},\ }\href@noop {} {\bibfield  {journal} {\bibinfo
  {journal} {Science}\ }\textbf {\bibinfo {volume} {372}},\ \bibinfo {pages}
  {259} (\bibinfo {year} {2021})}\BibitemShut {NoStop}%
\bibitem [{\citenamefont {Jamali}\ \emph {et~al.}(2014)\citenamefont {Jamali},
  \citenamefont {Gerhardt}, \citenamefont {Rezai}, \citenamefont {Frenner},
  \citenamefont {Fedder},\ and\ \citenamefont {Wrachtrup}}]{Jamali2014}%
  \BibitemOpen
  \bibfield  {author} {\bibinfo {author} {\bibfnamefont {M.}~\bibnamefont
  {Jamali}}, \bibinfo {author} {\bibfnamefont {I.}~\bibnamefont {Gerhardt}},
  \bibinfo {author} {\bibfnamefont {M.}~\bibnamefont {Rezai}}, \bibinfo
  {author} {\bibfnamefont {K.}~\bibnamefont {Frenner}}, \bibinfo {author}
  {\bibfnamefont {H.}~\bibnamefont {Fedder}},\ and\ \bibinfo {author}
  {\bibfnamefont {J.}~\bibnamefont {Wrachtrup}},\ }\bibfield  {title} {\bibinfo
  {title} {Microscopic diamond solid-immersion-lenses fabricated around single
  defect centers by focused ion beam milling},\ }\href@noop {} {\bibfield
  {journal} {\bibinfo  {journal} {Rev. Sci. Instrum.}\ }\textbf {\bibinfo
  {volume} {85}},\ \bibinfo {pages} {123703} (\bibinfo {year}
  {2014})}\BibitemShut {NoStop}%
\bibitem [{\citenamefont {Hadden}\ \emph {et~al.}(2010)\citenamefont {Hadden},
  \citenamefont {Harrison}, \citenamefont {Stanley-Clarke}, \citenamefont
  {Marseglia}, \citenamefont {Ho}, \citenamefont {Patton}, \citenamefont
  {O'Brien},\ and\ \citenamefont {Rarity}}]{Hadden2010}%
  \BibitemOpen
  \bibfield  {author} {\bibinfo {author} {\bibfnamefont {J.}~\bibnamefont
  {Hadden}}, \bibinfo {author} {\bibfnamefont {J.~P.}\ \bibnamefont
  {Harrison}}, \bibinfo {author} {\bibfnamefont {A.~C.}\ \bibnamefont
  {Stanley-Clarke}}, \bibinfo {author} {\bibfnamefont {L.}~\bibnamefont
  {Marseglia}}, \bibinfo {author} {\bibfnamefont {Y.-L.~D.}\ \bibnamefont
  {Ho}}, \bibinfo {author} {\bibfnamefont {B.~R.}\ \bibnamefont {Patton}},
  \bibinfo {author} {\bibfnamefont {J.~L.}\ \bibnamefont {O'Brien}},\ and\
  \bibinfo {author} {\bibfnamefont {J.~G.}\ \bibnamefont {Rarity}},\ }\bibfield
   {title} {\bibinfo {title} {Strongly enhanced photon collection from diamond
  defect centers under microfabricated integrated solid immersion lenses},\
  }\href@noop {} {\bibfield  {journal} {\bibinfo  {journal} {Appl. Phys.
  Lett.}\ }\textbf {\bibinfo {volume} {97}},\ \bibinfo {pages} {241901}
  (\bibinfo {year} {2010})}\BibitemShut {NoStop}%
\bibitem [{\citenamefont {Barbour}\ \emph {et~al.}(2011)\citenamefont
  {Barbour}, \citenamefont {Dalgarno}, \citenamefont {Curran}, \citenamefont
  {Nowak}, \citenamefont {Baker}, \citenamefont {Hall}, \citenamefont {Stoltz},
  \citenamefont {Petroff},\ and\ \citenamefont {Warburton}}]{Barbour2011b}%
  \BibitemOpen
  \bibfield  {author} {\bibinfo {author} {\bibfnamefont {R.~J.}\ \bibnamefont
  {Barbour}}, \bibinfo {author} {\bibfnamefont {P.~A.}\ \bibnamefont
  {Dalgarno}}, \bibinfo {author} {\bibfnamefont {A.}~\bibnamefont {Curran}},
  \bibinfo {author} {\bibfnamefont {K.~M.}\ \bibnamefont {Nowak}}, \bibinfo
  {author} {\bibfnamefont {H.~J.}\ \bibnamefont {Baker}}, \bibinfo {author}
  {\bibfnamefont {D.~R.}\ \bibnamefont {Hall}}, \bibinfo {author}
  {\bibfnamefont {N.~G.}\ \bibnamefont {Stoltz}}, \bibinfo {author}
  {\bibfnamefont {P.~M.}\ \bibnamefont {Petroff}},\ and\ \bibinfo {author}
  {\bibfnamefont {R.~J.}\ \bibnamefont {Warburton}},\ }\bibfield  {title}
  {\bibinfo {title} {A tunable microcavity},\ }\href@noop {} {\bibfield
  {journal} {\bibinfo  {journal} {J. Appl. Phys.}\ }\textbf {\bibinfo {volume}
  {110}},\ \bibinfo {pages} {053107} (\bibinfo {year} {2011})}\BibitemShut
  {NoStop}%
\bibitem [{\citenamefont {Albrecht}\ \emph {et~al.}(2013)\citenamefont
  {Albrecht}, \citenamefont {Bommer}, \citenamefont {Deutsch}, \citenamefont
  {Reichel},\ and\ \citenamefont {Becher}}]{Albrecht2013}%
  \BibitemOpen
  \bibfield  {author} {\bibinfo {author} {\bibfnamefont {R.}~\bibnamefont
  {Albrecht}}, \bibinfo {author} {\bibfnamefont {A.}~\bibnamefont {Bommer}},
  \bibinfo {author} {\bibfnamefont {C.}~\bibnamefont {Deutsch}}, \bibinfo
  {author} {\bibfnamefont {J.}~\bibnamefont {Reichel}},\ and\ \bibinfo {author}
  {\bibfnamefont {C.}~\bibnamefont {Becher}},\ }\bibfield  {title} {\bibinfo
  {title} {Coupling of a single nitrogen-vacancy center in diamond to a
  fiber-based microcavity},\ }\href@noop {} {\bibfield  {journal} {\bibinfo
  {journal} {Phys. Rev. Lett.}\ }\textbf {\bibinfo {volume} {110}},\ \bibinfo
  {pages} {243602} (\bibinfo {year} {2013})}\BibitemShut {NoStop}%
\bibitem [{\citenamefont {Johnson}\ \emph {et~al.}(2015)\citenamefont
  {Johnson}, \citenamefont {Dolan}, \citenamefont {Grange}, \citenamefont
  {Trichet}, \citenamefont {Hornecker}, \citenamefont {Chen}, \citenamefont
  {Weng}, \citenamefont {Hughes}, \citenamefont {Watt}, \citenamefont
  {Auff{\`{e}}ves},\ and\ \citenamefont {Smith}}]{Johnson2015}%
  \BibitemOpen
  \bibfield  {author} {\bibinfo {author} {\bibfnamefont {S.}~\bibnamefont
  {Johnson}}, \bibinfo {author} {\bibfnamefont {P.~R.}\ \bibnamefont {Dolan}},
  \bibinfo {author} {\bibfnamefont {T.}~\bibnamefont {Grange}}, \bibinfo
  {author} {\bibfnamefont {A.~A.~P.}\ \bibnamefont {Trichet}}, \bibinfo
  {author} {\bibfnamefont {G.}~\bibnamefont {Hornecker}}, \bibinfo {author}
  {\bibfnamefont {Y.~C.}\ \bibnamefont {Chen}}, \bibinfo {author}
  {\bibfnamefont {L.}~\bibnamefont {Weng}}, \bibinfo {author} {\bibfnamefont
  {G.~M.}\ \bibnamefont {Hughes}}, \bibinfo {author} {\bibfnamefont {A.~A.~R.}\
  \bibnamefont {Watt}}, \bibinfo {author} {\bibfnamefont {A.}~\bibnamefont
  {Auff{\`{e}}ves}},\ and\ \bibinfo {author} {\bibfnamefont {J.~M.}\
  \bibnamefont {Smith}},\ }\bibfield  {title} {\bibinfo {title} {Tunable cavity
  coupling of the zero phonon line of a nitrogen-vacancy defect in diamond},\
  }\href@noop {} {\bibfield  {journal} {\bibinfo  {journal} {New J. Phys.}\
  }\textbf {\bibinfo {volume} {17}},\ \bibinfo {pages} {122003} (\bibinfo
  {year} {2015})}\BibitemShut {NoStop}%
\bibitem [{\citenamefont {Kaupp}\ \emph {et~al.}(2016)\citenamefont {Kaupp},
  \citenamefont {H{\"{u}}mmer}, \citenamefont {Mader}, \citenamefont
  {Schlederer}, \citenamefont {Benedikter}, \citenamefont {Haeusser},
  \citenamefont {Chang}, \citenamefont {Fedder}, \citenamefont {H{\"{a}}nsch},\
  and\ \citenamefont {Hunger}}]{Kaupp2016}%
  \BibitemOpen
  \bibfield  {author} {\bibinfo {author} {\bibfnamefont {H.}~\bibnamefont
  {Kaupp}}, \bibinfo {author} {\bibfnamefont {T.}~\bibnamefont {H{\"{u}}mmer}},
  \bibinfo {author} {\bibfnamefont {M.}~\bibnamefont {Mader}}, \bibinfo
  {author} {\bibfnamefont {B.}~\bibnamefont {Schlederer}}, \bibinfo {author}
  {\bibfnamefont {J.}~\bibnamefont {Benedikter}}, \bibinfo {author}
  {\bibfnamefont {P.}~\bibnamefont {Haeusser}}, \bibinfo {author}
  {\bibfnamefont {H.~C.}\ \bibnamefont {Chang}}, \bibinfo {author}
  {\bibfnamefont {H.}~\bibnamefont {Fedder}}, \bibinfo {author} {\bibfnamefont
  {T.~W.}\ \bibnamefont {H{\"{a}}nsch}},\ and\ \bibinfo {author} {\bibfnamefont
  {D.}~\bibnamefont {Hunger}},\ }\bibfield  {title} {\bibinfo {title}
  {Purcell-enhanced single-photon emission from nitrogen-vacancy centers
  coupled to a tunable microcavity},\ }\href@noop {} {\bibfield  {journal}
  {\bibinfo  {journal} {Phys. Rev. Appl.}\ }\textbf {\bibinfo {volume} {6}},\
  \bibinfo {pages} {054010} (\bibinfo {year} {2016})}\BibitemShut {NoStop}%
\bibitem [{\citenamefont {Janitz}\ \emph {et~al.}(2015)\citenamefont {Janitz},
  \citenamefont {Ruf}, \citenamefont {Dimock}, \citenamefont {Bourassa},
  \citenamefont {Sankey},\ and\ \citenamefont {Childress}}]{Janitz2015}%
  \BibitemOpen
  \bibfield  {author} {\bibinfo {author} {\bibfnamefont {E.}~\bibnamefont
  {Janitz}}, \bibinfo {author} {\bibfnamefont {M.}~\bibnamefont {Ruf}},
  \bibinfo {author} {\bibfnamefont {M.}~\bibnamefont {Dimock}}, \bibinfo
  {author} {\bibfnamefont {A.}~\bibnamefont {Bourassa}}, \bibinfo {author}
  {\bibfnamefont {J.}~\bibnamefont {Sankey}},\ and\ \bibinfo {author}
  {\bibfnamefont {L.}~\bibnamefont {Childress}},\ }\bibfield  {title} {\bibinfo
  {title} {Fabry-perot microcavity for diamond-based photonics},\ }\href@noop
  {} {\bibfield  {journal} {\bibinfo  {journal} {Phys. Rev. A}\ }\textbf
  {\bibinfo {volume} {92}},\ \bibinfo {pages} {43844} (\bibinfo {year}
  {2015})}\BibitemShut {NoStop}%
\bibitem [{\citenamefont {Bogdanovi{\'{c}}}\ \emph {et~al.}(2017)\citenamefont
  {Bogdanovi{\'{c}}}, \citenamefont {van Dam}, \citenamefont {Bonato},
  \citenamefont {Coenen}, \citenamefont {Zwerver}, \citenamefont {Hensen},
  \citenamefont {Liddy}, \citenamefont {Fink}, \citenamefont {Reiserer},
  \citenamefont {Lon{\v{c}}ar},\ and\ \citenamefont {Hanson}}]{Bogdanovic2017}%
  \BibitemOpen
  \bibfield  {author} {\bibinfo {author} {\bibfnamefont {S.}~\bibnamefont
  {Bogdanovi{\'{c}}}}, \bibinfo {author} {\bibfnamefont {S.}~\bibnamefont {van
  Dam}}, \bibinfo {author} {\bibfnamefont {S.}~\bibnamefont {Bonato}}, \bibinfo
  {author} {\bibfnamefont {L.~C.}\ \bibnamefont {Coenen}}, \bibinfo {author}
  {\bibfnamefont {A.-M.~J.}\ \bibnamefont {Zwerver}}, \bibinfo {author}
  {\bibfnamefont {B.}~\bibnamefont {Hensen}}, \bibinfo {author} {\bibfnamefont
  {M.~S.~Z.}\ \bibnamefont {Liddy}}, \bibinfo {author} {\bibfnamefont
  {T.}~\bibnamefont {Fink}}, \bibinfo {author} {\bibfnamefont {A.}~\bibnamefont
  {Reiserer}}, \bibinfo {author} {\bibfnamefont {M.}~\bibnamefont
  {Lon{\v{c}}ar}},\ and\ \bibinfo {author} {\bibfnamefont {R.}~\bibnamefont
  {Hanson}},\ }\bibfield  {title} {\bibinfo {title} {Design and low-temperature
  characterization of a tunable microcavity for diamond-based quantum
  networks},\ }\href@noop {} {\bibfield  {journal} {\bibinfo  {journal} {Appl.
  Phys. Lett.}\ }\textbf {\bibinfo {volume} {110}},\ \bibinfo {pages} {171103}
  (\bibinfo {year} {2017})}\BibitemShut {NoStop}%
\bibitem [{\citenamefont {Riedel}\ \emph {et~al.}(2017)\citenamefont {Riedel},
  \citenamefont {S{\"{o}}llner}, \citenamefont {Shields}, \citenamefont
  {Starosielec}, \citenamefont {Appel}, \citenamefont {Neu}, \citenamefont
  {Maletinsky},\ and\ \citenamefont {Warburton}}]{Riedel2017c}%
  \BibitemOpen
  \bibfield  {author} {\bibinfo {author} {\bibfnamefont {D.}~\bibnamefont
  {Riedel}}, \bibinfo {author} {\bibfnamefont {I.}~\bibnamefont
  {S{\"{o}}llner}}, \bibinfo {author} {\bibfnamefont {B.~J.}\ \bibnamefont
  {Shields}}, \bibinfo {author} {\bibfnamefont {S.}~\bibnamefont
  {Starosielec}}, \bibinfo {author} {\bibfnamefont {P.}~\bibnamefont {Appel}},
  \bibinfo {author} {\bibfnamefont {E.}~\bibnamefont {Neu}}, \bibinfo {author}
  {\bibfnamefont {P.}~\bibnamefont {Maletinsky}},\ and\ \bibinfo {author}
  {\bibfnamefont {R.~J.}\ \bibnamefont {Warburton}},\ }\bibfield  {title}
  {\bibinfo {title} {Deterministic enhancement of coherent photon generation
  from a nitrogen-vacancy center in ultrapure diamond},\ }\href@noop {}
  {\bibfield  {journal} {\bibinfo  {journal} {Phys. Rev. X}\ }\textbf {\bibinfo
  {volume} {7}},\ \bibinfo {pages} {031040} (\bibinfo {year}
  {2017})}\BibitemShut {NoStop}%
\bibitem [{\citenamefont {Rohner}\ \emph {et~al.}(2019)\citenamefont {Rohner},
  \citenamefont {Happacher}, \citenamefont {Reiser}, \citenamefont {Tschudin},
  \citenamefont {Tallaire}, \citenamefont {Achard}, \citenamefont {Shields},\
  and\ \citenamefont {Maletinsky}}]{Rohner2019}%
  \BibitemOpen
  \bibfield  {author} {\bibinfo {author} {\bibfnamefont {D.}~\bibnamefont
  {Rohner}}, \bibinfo {author} {\bibfnamefont {J.}~\bibnamefont {Happacher}},
  \bibinfo {author} {\bibfnamefont {P.}~\bibnamefont {Reiser}}, \bibinfo
  {author} {\bibfnamefont {M.~A.}\ \bibnamefont {Tschudin}}, \bibinfo {author}
  {\bibfnamefont {A.}~\bibnamefont {Tallaire}}, \bibinfo {author}
  {\bibfnamefont {J.}~\bibnamefont {Achard}}, \bibinfo {author} {\bibfnamefont
  {B.~J.}\ \bibnamefont {Shields}},\ and\ \bibinfo {author} {\bibfnamefont
  {P.}~\bibnamefont {Maletinsky}},\ }\bibfield  {title} {\bibinfo {title}
  {(111)-oriented, single crystal diamond tips for nanoscale scanning probe
  imaging of out-of-plane magnetic fields},\ }\href@noop {} {\bibfield
  {journal} {\bibinfo  {journal} {Appl. Phys. Lett.}\ }\textbf {\bibinfo
  {volume} {115}},\ \bibinfo {pages} {192401} (\bibinfo {year}
  {2019})}\BibitemShut {NoStop}%
\bibitem [{\citenamefont {Hedrich}\ \emph {et~al.}(2020)\citenamefont
  {Hedrich}, \citenamefont {Rohner}, \citenamefont {Batzer}, \citenamefont
  {Maletinsky},\ and\ \citenamefont {Shields}}]{Hedrich2020}%
  \BibitemOpen
  \bibfield  {author} {\bibinfo {author} {\bibfnamefont {N.}~\bibnamefont
  {Hedrich}}, \bibinfo {author} {\bibfnamefont {D.}~\bibnamefont {Rohner}},
  \bibinfo {author} {\bibfnamefont {M.}~\bibnamefont {Batzer}}, \bibinfo
  {author} {\bibfnamefont {P.}~\bibnamefont {Maletinsky}},\ and\ \bibinfo
  {author} {\bibfnamefont {B.~J.}\ \bibnamefont {Shields}},\ }\bibfield
  {title} {\bibinfo {title} {Parabolic diamond scanning probes for single-spin
  magnetic field imaging},\ }\href@noop {} {\bibfield  {journal} {\bibinfo
  {journal} {Phys. Rev. Applied}\ }\textbf {\bibinfo {volume} {14}},\ \bibinfo
  {pages} {064007} (\bibinfo {year} {2020})}\BibitemShut {NoStop}%
\bibitem [{\citenamefont {Ruf}\ \emph {et~al.}(2019)\citenamefont {Ruf},
  \citenamefont {Ijspeert}, \citenamefont {{Van Dam}}, \citenamefont {{De
  Jong}}, \citenamefont {{Van Den Berg}}, \citenamefont {Evers},\ and\
  \citenamefont {Hanson}}]{Ruf2019a}%
  \BibitemOpen
  \bibfield  {author} {\bibinfo {author} {\bibfnamefont {M.}~\bibnamefont
  {Ruf}}, \bibinfo {author} {\bibfnamefont {M.}~\bibnamefont {Ijspeert}},
  \bibinfo {author} {\bibfnamefont {S.}~\bibnamefont {{Van Dam}}}, \bibinfo
  {author} {\bibfnamefont {N.}~\bibnamefont {{De Jong}}}, \bibinfo {author}
  {\bibfnamefont {H.}~\bibnamefont {{Van Den Berg}}}, \bibinfo {author}
  {\bibfnamefont {G.}~\bibnamefont {Evers}},\ and\ \bibinfo {author}
  {\bibfnamefont {R.}~\bibnamefont {Hanson}},\ }\bibfield  {title} {\bibinfo
  {title} {Optically coherent nitrogen-vacancy centers in micrometer-thin
  etched diamond membranes},\ }\href@noop {} {\bibfield  {journal} {\bibinfo
  {journal} {Nano Lett.}\ }\textbf {\bibinfo {volume} {19}},\ \bibinfo {pages}
  {3987} (\bibinfo {year} {2019})}\BibitemShut {NoStop}%
\bibitem [{\citenamefont {Tamarat}\ \emph {et~al.}(2006)\citenamefont
  {Tamarat}, \citenamefont {Gaebel}, \citenamefont {Rabeau}, \citenamefont
  {Khan}, \citenamefont {Greentree}, \citenamefont {Wilson}, \citenamefont
  {Hollenberg}, \citenamefont {Prawer}, \citenamefont {Hemmer}, \citenamefont
  {Jelezko},\ and\ \citenamefont {Wrachtrup}}]{Tamarat2006}%
  \BibitemOpen
  \bibfield  {author} {\bibinfo {author} {\bibfnamefont {P.}~\bibnamefont
  {Tamarat}}, \bibinfo {author} {\bibfnamefont {T.}~\bibnamefont {Gaebel}},
  \bibinfo {author} {\bibfnamefont {J.~R.}\ \bibnamefont {Rabeau}}, \bibinfo
  {author} {\bibfnamefont {M.}~\bibnamefont {Khan}}, \bibinfo {author}
  {\bibfnamefont {A.~D.}\ \bibnamefont {Greentree}}, \bibinfo {author}
  {\bibfnamefont {H.}~\bibnamefont {Wilson}}, \bibinfo {author} {\bibfnamefont
  {L.~C.~L.}\ \bibnamefont {Hollenberg}}, \bibinfo {author} {\bibfnamefont
  {S.}~\bibnamefont {Prawer}}, \bibinfo {author} {\bibfnamefont
  {P.}~\bibnamefont {Hemmer}}, \bibinfo {author} {\bibfnamefont
  {F.}~\bibnamefont {Jelezko}},\ and\ \bibinfo {author} {\bibfnamefont
  {J.}~\bibnamefont {Wrachtrup}},\ }\bibfield  {title} {\bibinfo {title} {Stark
  shift control of single optical centers in diamond},\ }\href@noop {}
  {\bibfield  {journal} {\bibinfo  {journal} {Phys. Rev. Lett.}\ }\textbf
  {\bibinfo {volume} {97}},\ \bibinfo {pages} {083002} (\bibinfo {year}
  {2006})}\BibitemShut {NoStop}%
\bibitem [{\citenamefont {Chakravarthi}\ \emph {et~al.}(2021)\citenamefont
  {Chakravarthi}, \citenamefont {Pederson}, \citenamefont {Kazi}, \citenamefont
  {Ivanov},\ and\ \citenamefont {Fu}}]{Chakravarthi2021}%
  \BibitemOpen
  \bibfield  {author} {\bibinfo {author} {\bibfnamefont {S.}~\bibnamefont
  {Chakravarthi}}, \bibinfo {author} {\bibfnamefont {C.}~\bibnamefont
  {Pederson}}, \bibinfo {author} {\bibfnamefont {Z.}~\bibnamefont {Kazi}},
  \bibinfo {author} {\bibfnamefont {A.}~\bibnamefont {Ivanov}},\ and\ \bibinfo
  {author} {\bibfnamefont {K.-M.}\ \bibnamefont {Fu}},\ }\bibfield  {title}
  {\bibinfo {title} {Impact of surface and laser-induced noise on the spectral
  stability of implanted nitrogen-vacancy centers in diamond},\ }\href@noop {}
  {\bibfield  {journal} {\bibinfo  {journal} {Phys. Rev. B}\ }\textbf {\bibinfo
  {volume} {104}},\ \bibinfo {pages} {085425} (\bibinfo {year}
  {2021})}\BibitemShut {NoStop}%
\bibitem [{\citenamefont {Chu}\ \emph {et~al.}(2014)\citenamefont {Chu},
  \citenamefont {{De Leon}}, \citenamefont {Shields}, \citenamefont {Hausmann},
  \citenamefont {Evans}, \citenamefont {Togan}, \citenamefont {Burek},
  \citenamefont {Markham}, \citenamefont {Stacey}, \citenamefont {Zibrov},
  \citenamefont {Yacoby}, \citenamefont {Twitchen}, \citenamefont
  {Lon{\v{c}}ar}, \citenamefont {Park}, \citenamefont {Maletinsky},\ and\
  \citenamefont {Lukin}}]{Chu2014a}%
  \BibitemOpen
  \bibfield  {author} {\bibinfo {author} {\bibfnamefont {Y.}~\bibnamefont
  {Chu}}, \bibinfo {author} {\bibfnamefont {N.~P.}\ \bibnamefont {{De Leon}}},
  \bibinfo {author} {\bibfnamefont {B.~J.}\ \bibnamefont {Shields}}, \bibinfo
  {author} {\bibfnamefont {B.}~\bibnamefont {Hausmann}}, \bibinfo {author}
  {\bibfnamefont {R.}~\bibnamefont {Evans}}, \bibinfo {author} {\bibfnamefont
  {E.}~\bibnamefont {Togan}}, \bibinfo {author} {\bibfnamefont {M.~J.}\
  \bibnamefont {Burek}}, \bibinfo {author} {\bibfnamefont {M.}~\bibnamefont
  {Markham}}, \bibinfo {author} {\bibfnamefont {A.}~\bibnamefont {Stacey}},
  \bibinfo {author} {\bibfnamefont {A.~S.}\ \bibnamefont {Zibrov}}, \bibinfo
  {author} {\bibfnamefont {A.}~\bibnamefont {Yacoby}}, \bibinfo {author}
  {\bibfnamefont {D.~J.}\ \bibnamefont {Twitchen}}, \bibinfo {author}
  {\bibfnamefont {M.}~\bibnamefont {Lon{\v{c}}ar}}, \bibinfo {author}
  {\bibfnamefont {H.}~\bibnamefont {Park}}, \bibinfo {author} {\bibfnamefont
  {P.}~\bibnamefont {Maletinsky}},\ and\ \bibinfo {author} {\bibfnamefont
  {M.~D.}\ \bibnamefont {Lukin}},\ }\bibfield  {title} {\bibinfo {title}
  {Coherent optical transitions in implanted nitrogen vacancy centers},\
  }\href@noop {} {\bibfield  {journal} {\bibinfo  {journal} {Nano Lett.}\
  }\textbf {\bibinfo {volume} {14}},\ \bibinfo {pages} {1982} (\bibinfo {year}
  {2014})}\BibitemShut {NoStop}%
\bibitem [{\citenamefont {Faraon}\ \emph {et~al.}(2012)\citenamefont {Faraon},
  \citenamefont {Santori}, \citenamefont {Huang}, \citenamefont {Acosta},\ and\
  \citenamefont {Beausoleil}}]{Faraon2012}%
  \BibitemOpen
  \bibfield  {author} {\bibinfo {author} {\bibfnamefont {A.}~\bibnamefont
  {Faraon}}, \bibinfo {author} {\bibfnamefont {C.}~\bibnamefont {Santori}},
  \bibinfo {author} {\bibfnamefont {Z.}~\bibnamefont {Huang}}, \bibinfo
  {author} {\bibfnamefont {V.~M.}\ \bibnamefont {Acosta}},\ and\ \bibinfo
  {author} {\bibfnamefont {R.~G.}\ \bibnamefont {Beausoleil}},\ }\bibfield
  {title} {\bibinfo {title} {Coupling of nitrogen-vacancy centers to photonic
  crystal cavities in monocrystalline diamond},\ }\href@noop {} {\bibfield
  {journal} {\bibinfo  {journal} {Phys. Rev. Lett.}\ }\textbf {\bibinfo
  {volume} {109}},\ \bibinfo {pages} {033604} (\bibinfo {year}
  {2012})}\BibitemShut {NoStop}%
\bibitem [{\citenamefont {Li}\ \emph {et~al.}(2015)\citenamefont {Li},
  \citenamefont {Schr{\"{o}}der}, \citenamefont {Chen}, \citenamefont {Walsh},
  \citenamefont {Bayn}, \citenamefont {Goldstein}, \citenamefont {Gaathon},
  \citenamefont {Trusheim}, \citenamefont {Lu}, \citenamefont {Mower},
  \citenamefont {Cotlet}, \citenamefont {Markham}, \citenamefont {Twitchen},\
  and\ \citenamefont {Englund}}]{Li2015}%
  \BibitemOpen
  \bibfield  {author} {\bibinfo {author} {\bibfnamefont {L.}~\bibnamefont
  {Li}}, \bibinfo {author} {\bibfnamefont {T.}~\bibnamefont {Schr{\"{o}}der}},
  \bibinfo {author} {\bibfnamefont {E.~H.}\ \bibnamefont {Chen}}, \bibinfo
  {author} {\bibfnamefont {M.}~\bibnamefont {Walsh}}, \bibinfo {author}
  {\bibfnamefont {I.}~\bibnamefont {Bayn}}, \bibinfo {author} {\bibfnamefont
  {J.}~\bibnamefont {Goldstein}}, \bibinfo {author} {\bibfnamefont
  {O.}~\bibnamefont {Gaathon}}, \bibinfo {author} {\bibfnamefont {M.~E.}\
  \bibnamefont {Trusheim}}, \bibinfo {author} {\bibfnamefont {M.}~\bibnamefont
  {Lu}}, \bibinfo {author} {\bibfnamefont {J.}~\bibnamefont {Mower}}, \bibinfo
  {author} {\bibfnamefont {M.}~\bibnamefont {Cotlet}}, \bibinfo {author}
  {\bibfnamefont {M.~L.}\ \bibnamefont {Markham}}, \bibinfo {author}
  {\bibfnamefont {D.~J.}\ \bibnamefont {Twitchen}},\ and\ \bibinfo {author}
  {\bibfnamefont {D.}~\bibnamefont {Englund}},\ }\bibfield  {title} {\bibinfo
  {title} {Coherent spin control of a nanocavity-enhanced qubit in diamond},\
  }\href@noop {} {\bibfield  {journal} {\bibinfo  {journal} {Nat. Commun.}\
  }\textbf {\bibinfo {volume} {6}},\ \bibinfo {pages} {1} (\bibinfo {year}
  {2015})}\BibitemShut {NoStop}%
\bibitem [{\citenamefont {Orphal-Kobin}\ \emph {et~al.}(2022)\citenamefont
  {Orphal-Kobin}, \citenamefont {Unterguggenberger}, \citenamefont
  {Pregnolato}, \citenamefont {Kemf}, \citenamefont {Matalla}, \citenamefont
  {Unger}, \citenamefont {Ostermay}, \citenamefont {Pieplow},\ and\
  \citenamefont {Schr{\"{o}}der}}]{Orphal-Kobin2022}%
  \BibitemOpen
  \bibfield  {author} {\bibinfo {author} {\bibfnamefont {L.}~\bibnamefont
  {Orphal-Kobin}}, \bibinfo {author} {\bibfnamefont {K.}~\bibnamefont
  {Unterguggenberger}}, \bibinfo {author} {\bibfnamefont {T.}~\bibnamefont
  {Pregnolato}}, \bibinfo {author} {\bibfnamefont {N.}~\bibnamefont {Kemf}},
  \bibinfo {author} {\bibfnamefont {M.}~\bibnamefont {Matalla}}, \bibinfo
  {author} {\bibfnamefont {R.-S.}\ \bibnamefont {Unger}}, \bibinfo {author}
  {\bibfnamefont {I.}~\bibnamefont {Ostermay}}, \bibinfo {author}
  {\bibfnamefont {G.}~\bibnamefont {Pieplow}},\ and\ \bibinfo {author}
  {\bibfnamefont {T.}~\bibnamefont {Schr{\"{o}}der}},\ }\bibfield  {title}
  {\bibinfo {title} {Optically coherent nitrogen-vacancy defect centers in
  diamond nanostructures},\ }\href@noop {} {\bibfield  {journal} {\bibinfo
  {journal} {arXiv:2203.05605v1}\ } (\bibinfo {year} {2022})}\BibitemShut
  {NoStop}%
\bibitem [{\citenamefont {{Van Dam}}\ \emph {et~al.}(2019)\citenamefont {{Van
  Dam}}, \citenamefont {Walsh}, \citenamefont {Degen}, \citenamefont {Bersin},
  \citenamefont {Mouradian}, \citenamefont {Galiullin}, \citenamefont {Ruf},
  \citenamefont {Ijspeert}, \citenamefont {Taminiau}, \citenamefont {Hanson},\
  and\ \citenamefont {Englund}}]{VanDam2019}%
  \BibitemOpen
  \bibfield  {author} {\bibinfo {author} {\bibfnamefont {S.~B.}\ \bibnamefont
  {{Van Dam}}}, \bibinfo {author} {\bibfnamefont {M.}~\bibnamefont {Walsh}},
  \bibinfo {author} {\bibfnamefont {M.~J.}\ \bibnamefont {Degen}}, \bibinfo
  {author} {\bibfnamefont {E.}~\bibnamefont {Bersin}}, \bibinfo {author}
  {\bibfnamefont {S.~L.}\ \bibnamefont {Mouradian}}, \bibinfo {author}
  {\bibfnamefont {A.}~\bibnamefont {Galiullin}}, \bibinfo {author}
  {\bibfnamefont {M.}~\bibnamefont {Ruf}}, \bibinfo {author} {\bibfnamefont
  {M.}~\bibnamefont {Ijspeert}}, \bibinfo {author} {\bibfnamefont {T.~H.}\
  \bibnamefont {Taminiau}}, \bibinfo {author} {\bibfnamefont {R.}~\bibnamefont
  {Hanson}},\ and\ \bibinfo {author} {\bibfnamefont {D.~R.}\ \bibnamefont
  {Englund}},\ }\bibfield  {title} {\bibinfo {title} {Optical coherence of
  diamond nitrogen-vacancy centers formed by ion implantation and annealing},\
  }\href@noop {} {\bibfield  {journal} {\bibinfo  {journal} {Phys. Rev. B}\
  }\textbf {\bibinfo {volume} {99}},\ \bibinfo {pages} {161203} (\bibinfo
  {year} {2019})}\BibitemShut {NoStop}%
\bibitem [{\citenamefont {Kasperczyk}\ \emph {et~al.}(2020)\citenamefont
  {Kasperczyk}, \citenamefont {Zuber}, \citenamefont {K{\"o}lbl}, \citenamefont
  {Yurgens}, \citenamefont {Fl{\aa}gan}, \citenamefont {Jakubczyk},
  \citenamefont {Shields}, \citenamefont {Warburton},\ and\ \citenamefont
  {Maletinsky}}]{Kasperczyk2020}%
  \BibitemOpen
  \bibfield  {author} {\bibinfo {author} {\bibfnamefont {M.}~\bibnamefont
  {Kasperczyk}}, \bibinfo {author} {\bibfnamefont {J.~A.}\ \bibnamefont
  {Zuber}}, \bibinfo {author} {\bibfnamefont {J.}~\bibnamefont {K{\"o}lbl}},
  \bibinfo {author} {\bibfnamefont {V.}~\bibnamefont {Yurgens}}, \bibinfo
  {author} {\bibfnamefont {S.}~\bibnamefont {Fl{\aa}gan}}, \bibinfo {author}
  {\bibfnamefont {T.}~\bibnamefont {Jakubczyk}}, \bibinfo {author}
  {\bibfnamefont {B.~J.}\ \bibnamefont {Shields}}, \bibinfo {author}
  {\bibfnamefont {R.~J.}\ \bibnamefont {Warburton}},\ and\ \bibinfo {author}
  {\bibfnamefont {P.}~\bibnamefont {Maletinsky}},\ }\bibfield  {title}
  {\bibinfo {title} {Statistically modeling optical linewidths of nitrogen
  vacancy centers in microstructures},\ }\href@noop {} {\bibfield  {journal}
  {\bibinfo  {journal} {Phys. Rev. B}\ }\textbf {\bibinfo {volume} {102}},\
  \bibinfo {pages} {075312} (\bibinfo {year} {2020})}\BibitemShut {NoStop}%
\bibitem [{\citenamefont {Chen}\ \emph {et~al.}(2017)\citenamefont {Chen},
  \citenamefont {Salter}, \citenamefont {Knauer}, \citenamefont {Weng},
  \citenamefont {Frangeskou}, \citenamefont {Stephen}, \citenamefont {Ishmael},
  \citenamefont {Dolan}, \citenamefont {Johnson}, \citenamefont {Green},
  \citenamefont {Morley}, \citenamefont {Newton}, \citenamefont {Rarity},
  \citenamefont {Booth},\ and\ \citenamefont {Smith}}]{Chen2017b}%
  \BibitemOpen
  \bibfield  {author} {\bibinfo {author} {\bibfnamefont {Y.-C.}\ \bibnamefont
  {Chen}}, \bibinfo {author} {\bibfnamefont {P.~S.}\ \bibnamefont {Salter}},
  \bibinfo {author} {\bibfnamefont {S.}~\bibnamefont {Knauer}}, \bibinfo
  {author} {\bibfnamefont {L.}~\bibnamefont {Weng}}, \bibinfo {author}
  {\bibfnamefont {A.~C.}\ \bibnamefont {Frangeskou}}, \bibinfo {author}
  {\bibfnamefont {C.~J.}\ \bibnamefont {Stephen}}, \bibinfo {author}
  {\bibfnamefont {S.~N.}\ \bibnamefont {Ishmael}}, \bibinfo {author}
  {\bibfnamefont {P.~R.}\ \bibnamefont {Dolan}}, \bibinfo {author}
  {\bibfnamefont {S.}~\bibnamefont {Johnson}}, \bibinfo {author} {\bibfnamefont
  {B.~L.}\ \bibnamefont {Green}}, \bibinfo {author} {\bibfnamefont {G.~W.}\
  \bibnamefont {Morley}}, \bibinfo {author} {\bibfnamefont {M.~E.}\
  \bibnamefont {Newton}}, \bibinfo {author} {\bibfnamefont {J.~G.}\
  \bibnamefont {Rarity}}, \bibinfo {author} {\bibfnamefont {M.~J.}\
  \bibnamefont {Booth}},\ and\ \bibinfo {author} {\bibfnamefont {J.~M.}\
  \bibnamefont {Smith}},\ }\bibfield  {title} {\bibinfo {title} {Laser writing
  of coherent colour centres in diamond},\ }\href@noop {} {\bibfield  {journal}
  {\bibinfo  {journal} {Nat. Photonics}\ }\textbf {\bibinfo {volume} {11}},\
  \bibinfo {pages} {77} (\bibinfo {year} {2017})}\BibitemShut {NoStop}%
\bibitem [{\citenamefont {Yurgens}\ \emph {et~al.}(2021)\citenamefont
  {Yurgens}, \citenamefont {Zuber}, \citenamefont {Fl{\aa}gan}, \citenamefont
  {De~Luca}, \citenamefont {Shields}, \citenamefont {Zardo}, \citenamefont
  {Maletinsky}, \citenamefont {Warburton},\ and\ \citenamefont
  {Jakubczyk}}]{Yurgens2021}%
  \BibitemOpen
  \bibfield  {author} {\bibinfo {author} {\bibfnamefont {V.}~\bibnamefont
  {Yurgens}}, \bibinfo {author} {\bibfnamefont {J.~A.}\ \bibnamefont {Zuber}},
  \bibinfo {author} {\bibfnamefont {S.}~\bibnamefont {Fl{\aa}gan}}, \bibinfo
  {author} {\bibfnamefont {M.}~\bibnamefont {De~Luca}}, \bibinfo {author}
  {\bibfnamefont {B.~J.}\ \bibnamefont {Shields}}, \bibinfo {author}
  {\bibfnamefont {I.}~\bibnamefont {Zardo}}, \bibinfo {author} {\bibfnamefont
  {P.}~\bibnamefont {Maletinsky}}, \bibinfo {author} {\bibfnamefont {R.~J.}\
  \bibnamefont {Warburton}},\ and\ \bibinfo {author} {\bibfnamefont
  {T.}~\bibnamefont {Jakubczyk}},\ }\bibfield  {title} {\bibinfo {title}
  {Low-charge-noise nitrogen-vacancy centers in diamond created using laser
  writing with a solid-immersion lens},\ }\href@noop {} {\bibfield  {journal}
  {\bibinfo  {journal} {ACS Photonics}\ }\textbf {\bibinfo {volume} {8}},\
  \bibinfo {pages} {1726–1734} (\bibinfo {year} {2021})}\BibitemShut
  {NoStop}%
\bibitem [{\citenamefont {Lekavicius}\ \emph {et~al.}(2019)\citenamefont
  {Lekavicius}, \citenamefont {Oo},\ and\ \citenamefont
  {Wang}}]{Lekavicius2019}%
  \BibitemOpen
  \bibfield  {author} {\bibinfo {author} {\bibfnamefont {I.}~\bibnamefont
  {Lekavicius}}, \bibinfo {author} {\bibfnamefont {T.}~\bibnamefont {Oo}},\
  and\ \bibinfo {author} {\bibfnamefont {H.}~\bibnamefont {Wang}},\ }\bibfield
  {title} {\bibinfo {title} {Diamond {L}amb wave spin-mechanical resonators
  with optically coherent nitrogen vacancy centers},\ }\href@noop {} {\bibfield
   {journal} {\bibinfo  {journal} {J. Appl. Phys.}\ }\textbf {\bibinfo {volume}
  {126}},\ \bibinfo {pages} {214301} (\bibinfo {year} {2019})}\BibitemShut
  {NoStop}%
\bibitem [{\citenamefont {Naydenov}\ \emph {et~al.}(2010)\citenamefont
  {Naydenov}, \citenamefont {Richter}, \citenamefont {Beck}, \citenamefont
  {Steiner}, \citenamefont {Neumann}, \citenamefont {Balasubramanian},
  \citenamefont {Achard}, \citenamefont {Jelezko}, \citenamefont {Wrachtrup},\
  and\ \citenamefont {Kalish}}]{Naydenov2010}%
  \BibitemOpen
  \bibfield  {author} {\bibinfo {author} {\bibfnamefont {B.}~\bibnamefont
  {Naydenov}}, \bibinfo {author} {\bibfnamefont {V.}~\bibnamefont {Richter}},
  \bibinfo {author} {\bibfnamefont {J.}~\bibnamefont {Beck}}, \bibinfo {author}
  {\bibfnamefont {M.}~\bibnamefont {Steiner}}, \bibinfo {author} {\bibfnamefont
  {P.}~\bibnamefont {Neumann}}, \bibinfo {author} {\bibfnamefont
  {G.}~\bibnamefont {Balasubramanian}}, \bibinfo {author} {\bibfnamefont
  {J.}~\bibnamefont {Achard}}, \bibinfo {author} {\bibfnamefont
  {F.}~\bibnamefont {Jelezko}}, \bibinfo {author} {\bibfnamefont
  {J.}~\bibnamefont {Wrachtrup}},\ and\ \bibinfo {author} {\bibfnamefont
  {R.}~\bibnamefont {Kalish}},\ }\bibfield  {title} {\bibinfo {title} {Enhanced
  generation of single optically active spins in diamond by ion implantation},\
  }\href@noop {} {\bibfield  {journal} {\bibinfo  {journal} {Appl. Phys.
  Lett.}\ }\textbf {\bibinfo {volume} {96}},\ \bibinfo {pages} {163108}
  (\bibinfo {year} {2010})}\BibitemShut {NoStop}%
\bibitem [{\citenamefont {Schwartz}\ \emph {et~al.}(2012)\citenamefont
  {Schwartz}, \citenamefont {Aloni}, \citenamefont {Ogletree},\ and\
  \citenamefont {Schenkel}}]{Schwartz2012}%
  \BibitemOpen
  \bibfield  {author} {\bibinfo {author} {\bibfnamefont {J.}~\bibnamefont
  {Schwartz}}, \bibinfo {author} {\bibfnamefont {S.}~\bibnamefont {Aloni}},
  \bibinfo {author} {\bibfnamefont {D.~F.}\ \bibnamefont {Ogletree}},\ and\
  \bibinfo {author} {\bibfnamefont {T.}~\bibnamefont {Schenkel}},\ }\bibfield
  {title} {\bibinfo {title} {Effects of low-energy electron irradiation on
  formation of nitrogen--vacancy centers in single-crystal diamond},\
  }\href@noop {} {\bibfield  {journal} {\bibinfo  {journal} {New J. Phys.}\
  }\textbf {\bibinfo {volume} {14}},\ \bibinfo {pages} {043024} (\bibinfo
  {year} {2012})}\BibitemShut {NoStop}%
\bibitem [{\citenamefont {Kim}\ \emph {et~al.}(2012)\citenamefont {Kim},
  \citenamefont {Acosta}, \citenamefont {Bauch}, \citenamefont {Budker},\ and\
  \citenamefont {Hemmer}}]{Kim2012}%
  \BibitemOpen
  \bibfield  {author} {\bibinfo {author} {\bibfnamefont {E.}~\bibnamefont
  {Kim}}, \bibinfo {author} {\bibfnamefont {V.~M.}\ \bibnamefont {Acosta}},
  \bibinfo {author} {\bibfnamefont {E.}~\bibnamefont {Bauch}}, \bibinfo
  {author} {\bibfnamefont {D.}~\bibnamefont {Budker}},\ and\ \bibinfo {author}
  {\bibfnamefont {P.~R.}\ \bibnamefont {Hemmer}},\ }\bibfield  {title}
  {\bibinfo {title} {Electron spin resonance shift and linewidth broadening of
  nitrogen-vacancy centers in diamond as a function of electron irradiation
  dose},\ }\href@noop {} {\bibfield  {journal} {\bibinfo  {journal} {Appl.
  Phys. Lett.}\ }\textbf {\bibinfo {volume} {101}},\ \bibinfo {pages} {082410}
  (\bibinfo {year} {2012})}\BibitemShut {NoStop}%
\bibitem [{\citenamefont {Appel}\ \emph {et~al.}(2016)\citenamefont {Appel},
  \citenamefont {Neu}, \citenamefont {Ganzhorn}, \citenamefont {Barfuss},
  \citenamefont {Batzer}, \citenamefont {Gratz}, \citenamefont
  {Tsch{\"{o}}pe},\ and\ \citenamefont {Maletinsky}}]{Appel2016}%
  \BibitemOpen
  \bibfield  {author} {\bibinfo {author} {\bibfnamefont {P.}~\bibnamefont
  {Appel}}, \bibinfo {author} {\bibfnamefont {E.}~\bibnamefont {Neu}}, \bibinfo
  {author} {\bibfnamefont {E.}~\bibnamefont {Ganzhorn}}, \bibinfo {author}
  {\bibfnamefont {A.}~\bibnamefont {Barfuss}}, \bibinfo {author} {\bibfnamefont
  {M.}~\bibnamefont {Batzer}}, \bibinfo {author} {\bibfnamefont
  {M.}~\bibnamefont {Gratz}}, \bibinfo {author} {\bibfnamefont
  {A.}~\bibnamefont {Tsch{\"{o}}pe}},\ and\ \bibinfo {author} {\bibfnamefont
  {P.}~\bibnamefont {Maletinsky}},\ }\bibfield  {title} {\bibinfo {title}
  {Fabrication of all diamond scanning probes for nanoscale magnetometry},\
  }\href@noop {} {\bibfield  {journal} {\bibinfo  {journal} {Rev. Sci.
  Instrum.}\ }\textbf {\bibinfo {volume} {87}},\ \bibinfo {pages} {063703}
  (\bibinfo {year} {2016})}\BibitemShut {NoStop}%
\bibitem [{\citenamefont {Challier}\ \emph {et~al.}(2018)\citenamefont
  {Challier}, \citenamefont {Sonusen}, \citenamefont {Barfuss}, \citenamefont
  {Rohner}, \citenamefont {Riedel}, \citenamefont {Koelbl}, \citenamefont
  {Ganzhorn}, \citenamefont {Appel}, \citenamefont {Maletinsky},\ and\
  \citenamefont {Neu}}]{Challier2018}%
  \BibitemOpen
  \bibfield  {author} {\bibinfo {author} {\bibfnamefont {M.}~\bibnamefont
  {Challier}}, \bibinfo {author} {\bibfnamefont {S.}~\bibnamefont {Sonusen}},
  \bibinfo {author} {\bibfnamefont {A.}~\bibnamefont {Barfuss}}, \bibinfo
  {author} {\bibfnamefont {D.}~\bibnamefont {Rohner}}, \bibinfo {author}
  {\bibfnamefont {D.}~\bibnamefont {Riedel}}, \bibinfo {author} {\bibfnamefont
  {J.}~\bibnamefont {Koelbl}}, \bibinfo {author} {\bibfnamefont
  {M.}~\bibnamefont {Ganzhorn}}, \bibinfo {author} {\bibfnamefont
  {P.}~\bibnamefont {Appel}}, \bibinfo {author} {\bibfnamefont
  {P.}~\bibnamefont {Maletinsky}},\ and\ \bibinfo {author} {\bibfnamefont
  {E.}~\bibnamefont {Neu}},\ }\bibfield  {title} {\bibinfo {title} {Advanced
  fabrication of single-crystal diamond membranes for quantum technologies},\
  }\href@noop {} {\bibfield  {journal} {\bibinfo  {journal} {Micromachines}\
  }\textbf {\bibinfo {volume} {9}},\ \bibinfo {pages} {148} (\bibinfo {year}
  {2018})}\BibitemShut {NoStop}%
\bibitem [{\citenamefont {Barfuss}\ \emph {et~al.}(2015)\citenamefont
  {Barfuss}, \citenamefont {Teissier}, \citenamefont {Neu}, \citenamefont
  {Nunnenkamp},\ and\ \citenamefont {Maletinsky}}]{Barfuss2015}%
  \BibitemOpen
  \bibfield  {author} {\bibinfo {author} {\bibfnamefont {A.}~\bibnamefont
  {Barfuss}}, \bibinfo {author} {\bibfnamefont {J.}~\bibnamefont {Teissier}},
  \bibinfo {author} {\bibfnamefont {E.}~\bibnamefont {Neu}}, \bibinfo {author}
  {\bibfnamefont {A.}~\bibnamefont {Nunnenkamp}},\ and\ \bibinfo {author}
  {\bibfnamefont {P.}~\bibnamefont {Maletinsky}},\ }\bibfield  {title}
  {\bibinfo {title} {Strong mechanical driving of a single electron spin},\
  }\href@noop {} {\bibfield  {journal} {\bibinfo  {journal} {Nat. Phys.}\
  }\textbf {\bibinfo {volume} {11}},\ \bibinfo {pages} {820} (\bibinfo {year}
  {2015})}\BibitemShut {NoStop}%
\bibitem [{\citenamefont {Bernien}\ \emph {et~al.}(2013)\citenamefont
  {Bernien}, \citenamefont {Hensen}, \citenamefont {Pfaff}, \citenamefont
  {Koolstra}, \citenamefont {Blok}, \citenamefont {Robledo}, \citenamefont
  {Taminiau}, \citenamefont {Markham}, \citenamefont {Twitchen}, \citenamefont
  {Childress},\ and\ \citenamefont {Hanson}}]{Bernien2013}%
  \BibitemOpen
  \bibfield  {author} {\bibinfo {author} {\bibfnamefont {H.}~\bibnamefont
  {Bernien}}, \bibinfo {author} {\bibfnamefont {B.}~\bibnamefont {Hensen}},
  \bibinfo {author} {\bibfnamefont {W.}~\bibnamefont {Pfaff}}, \bibinfo
  {author} {\bibfnamefont {G.}~\bibnamefont {Koolstra}}, \bibinfo {author}
  {\bibfnamefont {M.~S.}\ \bibnamefont {Blok}}, \bibinfo {author}
  {\bibfnamefont {L.}~\bibnamefont {Robledo}}, \bibinfo {author} {\bibfnamefont
  {T.~H.}\ \bibnamefont {Taminiau}}, \bibinfo {author} {\bibfnamefont
  {M.}~\bibnamefont {Markham}}, \bibinfo {author} {\bibfnamefont {D.~J.}\
  \bibnamefont {Twitchen}}, \bibinfo {author} {\bibfnamefont {L.}~\bibnamefont
  {Childress}},\ and\ \bibinfo {author} {\bibfnamefont {R.}~\bibnamefont
  {Hanson}},\ }\bibfield  {title} {\bibinfo {title} {Heralded entanglement
  between solid-state qubits separated by three metres},\ }\href@noop {}
  {\bibfield  {journal} {\bibinfo  {journal} {Nature}\ }\textbf {\bibinfo
  {volume} {497}},\ \bibinfo {pages} {86} (\bibinfo {year} {2013})}\BibitemShut
  {NoStop}%
\bibitem [{\citenamefont {Legero}\ \emph {et~al.}(2003)\citenamefont {Legero},
  \citenamefont {Wilk}, \citenamefont {Kuhn},\ and\ \citenamefont
  {Rempe}}]{Legero2003}%
  \BibitemOpen
  \bibfield  {author} {\bibinfo {author} {\bibfnamefont {T.}~\bibnamefont
  {Legero}}, \bibinfo {author} {\bibfnamefont {T.}~\bibnamefont {Wilk}},
  \bibinfo {author} {\bibfnamefont {A.}~\bibnamefont {Kuhn}},\ and\ \bibinfo
  {author} {\bibfnamefont {G.}~\bibnamefont {Rempe}},\ }\bibfield  {title}
  {\bibinfo {title} {Time-resolved two-photon quantum interference},\
  }\href@noop {} {\bibfield  {journal} {\bibinfo  {journal} {Appl. Phys. B.}\
  }\textbf {\bibinfo {volume} {77}},\ \bibinfo {pages} {797} (\bibinfo {year}
  {2003})}\BibitemShut {NoStop}%
\bibitem [{\citenamefont {Legero}\ \emph {et~al.}(2006)\citenamefont {Legero},
  \citenamefont {Wilk}, \citenamefont {Kuhn},\ and\ \citenamefont
  {Rempe}}]{Legero2006}%
  \BibitemOpen
  \bibfield  {author} {\bibinfo {author} {\bibfnamefont {T.}~\bibnamefont
  {Legero}}, \bibinfo {author} {\bibfnamefont {T.}~\bibnamefont {Wilk}},
  \bibinfo {author} {\bibfnamefont {A.}~\bibnamefont {Kuhn}},\ and\ \bibinfo
  {author} {\bibfnamefont {G.}~\bibnamefont {Rempe}},\ }\bibfield  {title}
  {\bibinfo {title} {Characterization of single photons using two-photon
  interference},\ }\href@noop {} {\bibfield  {journal} {\bibinfo  {journal}
  {Adv. At. Mol. Opt. Phys.}\ }\textbf {\bibinfo {volume} {53}},\ \bibinfo
  {pages} {253} (\bibinfo {year} {2006})}\BibitemShut {NoStop}%
\bibitem [{\citenamefont {Kambs}\ and\ \citenamefont
  {Becher}(2018)}]{Kambs2018}%
  \BibitemOpen
  \bibfield  {author} {\bibinfo {author} {\bibfnamefont {B.}~\bibnamefont
  {Kambs}}\ and\ \bibinfo {author} {\bibfnamefont {C.}~\bibnamefont {Becher}},\
  }\bibfield  {title} {\bibinfo {title} {Limitations on the
  indistinguishability of photons from remote solid state sources},\
  }\href@noop {} {\bibfield  {journal} {\bibinfo  {journal} {New J. Phys.}\
  }\textbf {\bibinfo {volume} {20}},\ \bibinfo {pages} {115003} (\bibinfo
  {year} {2018})}\BibitemShut {NoStop}%
\bibitem [{\citenamefont {Barrett}\ and\ \citenamefont
  {Kok}(2005)}]{Barrett2005}%
  \BibitemOpen
  \bibfield  {author} {\bibinfo {author} {\bibfnamefont {S.~D.}\ \bibnamefont
  {Barrett}}\ and\ \bibinfo {author} {\bibfnamefont {P.}~\bibnamefont {Kok}},\
  }\bibfield  {title} {\bibinfo {title} {Efficient high-fidelity quantum
  computation using matter qubits and linear optics},\ }\href@noop {}
  {\bibfield  {journal} {\bibinfo  {journal} {Phys. Rev. A}\ }\textbf {\bibinfo
  {volume} {71}},\ \bibinfo {pages} {060310} (\bibinfo {year}
  {2005})}\BibitemShut {NoStop}%
\end{thebibliography}%
\end{document}